\documentclass{SciPost}

\hypersetup{
    colorlinks,
    linkcolor={red!50!black},
    citecolor={blue!50!black},
    urlcolor={blue!80!black}
}

\usepackage[bitstream-charter]{mathdesign}
\DeclareSymbolFont{usualmathcal}{OMS}{cmsy}{m}{n}
\DeclareSymbolFontAlphabet{\mathcal}{usualmathcal}

\fancypagestyle{SPstyle}{
\fancyhf{}
\lhead{\colorbox{scipostblue}{\bf \color{white} ~SciPost Physics }}
\rhead{{\bf \color{scipostdeepblue} ~Submission }}

\fancyfoot[C]{\textbf{\thepage}}
}

\usepackage{braket}
\usepackage{bm}
\usepackage{textcase}
\usepackage{subcaption}
\usepackage{listings}
\usepackage{bbm}
\usepackage[capitalise]{cleveref}
\usepackage{tikz}
\usepackage[section]{placeins} 

\renewcommand{\d}{\ensuremath{\mathrm{d}}}
\newcommand{\M}{\mathcal{M}}

\renewcommand{\Re}{\mathrm{Re}}
\renewcommand{\Im}{\mathrm{Im}}
\newcommand{\R}{\mathbb{R}}
\newcommand{\Id}{\mathbbm{1}}
\newcommand{\Tr}{\mathrm{Tr}}

\newcommand{\adj}{\mathrm{adj}}

\renewcommand{\H}{\hat{H}}
\renewcommand{\O}{\mathcal{O}}
\newcommand{\h}{\bm h}
\newcommand{\hhsub}{\bm f_\times}

\newcommand{\coord}{\alpha}
\newcommand{\ccoord}{\bm{\alpha}}
\newcommand{\dpind}{q}
\newcommand{\CC}{\mathcal C}
\newcommand{\PT}{\mathcal{PT}}

\definecolor{MyBlue}{rgb}{0, 0.4, 1}

\begin{document}

\pagestyle{SPstyle}

\begin{center}{\Large \textbf{\color{scipostdeepblue}{
Quantum geometry of connected state manifolds: When diabolic points act as bridges between eigenstate manifolds
}}}\end{center}

\begin{center}\textbf{
Jan St\v{r}ele\v{c}ek$^{\ast}$, Jakub Novotn\'{y}$^{\dagger}$ and Pavel Cejnar\textsuperscript{}
}\end{center}

\begin{center}
Institute of Particle and Nuclear Physics, Faculty of Mathematics and Physics, Charles University, V Holešovičkách 2, 18000 Prague 8, Czech Republic
\\[\baselineskip]
$\star$ \href{mailto:email1}{\small jan.strelecek@matfyz.cuni.cz}\,,\quad
$\dagger$ \href{mailto:email2}{\small jakub.novotny@matfyz.cuni.cz}
\end{center}

\section*{\color{scipostdeepblue}{Abstract}}
\textbf{\boldmath{%
Parametric Hamiltonians often exhibit point-like spectral degeneracies, called diabolic points (DPs) or conical intersections, which can lead to singularities in the Provost--Vallee metric of eigenstate manifolds. We regularise the metric by a~transformation to polar coordinates and develop a~formalism in which DPs act as bridges between adjacent eigenstate manifolds, glueing them into a~single connected state manifold (CSM). We characterise the topology of this structure and extend the local rules for nodal lines governing the Berry phase to global ones. On the CSM, the class of geodesics can be enlarged by those passing through the DPs. They can be stable with respect to their initial direction, and can be shorter than those constrained to a~single eigenstate manifold, providing new geodesic shortcuts. Further, the CSM provides a~new mechanism for Berry phase computation along paths traversing the DPs.
}}

\vspace{\baselineskip}



\vspace{10pt}
\noindent\rule{\textwidth}{1pt}
\tableofcontents
\noindent\rule{\textwidth}{1pt}
\vspace{10pt}

\section{Introduction}
\label{sec:introduction}
The quantum geometric tensor (QGT) is an important tool for studying parametric quantum systems. It encodes both the Provost--Vallee (Fubini--Study) metric~\cite{Provost80,Wootters81,Anandan90,Graf21}, which measures infinitesimal distances between quantum states, and the Berry curvature~\cite{Simon83, Berry84, Cheng13}, which governs geometric phases. The Provost--Vallee metric equips each eigenstate manifold with a~positive semidefinite metric structure, allowing us to study how distances, geodesics~\cite{Kumar12, Tomka16, Bleu18}, and curvature in parametric space relate to other physical quantities. 

The geometric structure appears across many fields. In condensed matter it describes phase transitions~\cite{Venuti07, Kumar12, Kumar14, Zanardi07}, topological insulators~\cite{Resta11, Mera21, Guillot25}, superfluidity~\cite{Peotta15, Torma22}, and electronic transport~\cite{Xiao10}. In quantum information it underlies error mitigation~\cite{Wierichs20, Atif22, Haddou25, Halla25}, sensing~\cite{Zeng24} and quantum state driving~\cite{Cejnar23, Matus23, Matus25}. The QGT has also been studied in various setups developing the methodology itself~\cite{Carollo20,Liska21,Hetenyi23,Dey12,Henriet18,Jaiswal21,Gutierrez21,Gutierrez22} and measured directly in experiments~\cite{Kolodrubetz13, Lin23, Yi23, Mingu24}.

The Provost--Vallee metric quantifies the sensitivity of eigenvectors to parameter variations in the Hamiltonian: the faster the eigenvectors vary with parameters, the larger the metric. This sensitivity is strongly enhanced when the spectral gap between nearby levels becomes small, and the metric typically diverges as a~gap closes. If these closures happen at isolated degeneracies, where the energy surfaces meet in a~characteristic double-cone structure, they are called diabolic points (DPs) or conical intersections~\cite{Berry84Diabolical,Yarkony96,Worth04}. DPs are ubiquitous: they produce quantised Berry phases under adiabatic encirclement \cite{Berry84, Louvet23}, they open non-radiative relaxation pathways in molecular photochemistry~\cite{Levine07, Matsika11, Malhado14}, and have been simulated on quantum computers~\cite{Wang24, Zhao24QC}. 

DPs pose fundamental difficulties for the geometric description. At the DP, the eigenstate is not uniquely defined, the metric components usually diverge, and individual eigenstate manifolds are treated as disconnected spaces bounded by these degeneracies. As a~consequence, geodesics terminate at the DPs \cite{Strelecek25}, and calculating the Berry phase for paths passing through DPs requires special gauge constructions~\cite{Ju25}. 

\begin{figure}
    \centering
    \includegraphics{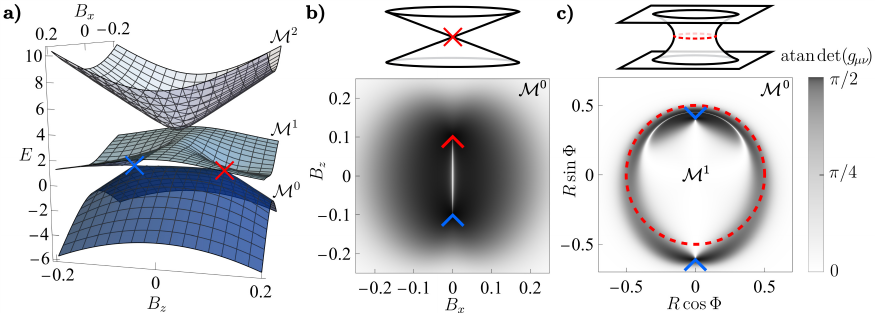}
    \caption{Illustration of the bridge connecting ground and first excited state manifolds of a~spin-1 model in a~magnetic field $\bm B = (B_x, 0, B_z)$, see Sec.~\ref{sec:spin-1-system}. (a) 3D plot of spectrum in a~$(B_x, B_z)$ parametric space. Red and blue crosses mark DPs between ground and first excited states, creating CSM from two eigenstate manifolds. The second state manifold is disconnected from it due to a~finite energy gap. (b) Determinant of the ground state metric tensor $g_{\mu\nu}^{(0)}$ in Cartesian coordinates manifests divergences near DPs, with a~schematic illustration of one of the energy crossings on top. (c) Regularised metric tensor in \emph{stretched coordinates} $(R,\Phi)$ around one of the DPs (red cross with positive $B_z$). The DP is transformed into a~boundary (red, dashed) between the ground state $\M^0$ (outside), and the first excited state manifold $\M^1$ (inside). This demonstrates how each of the two DPs forms a~bridge between eigenstate manifolds, as sketched in panel (c). Blue and red halves of crosses mark DPs as seen from the manifold upon which they are drawn.}
    \label{fig:illustration}
\end{figure}

To tackle these problems, we build upon the formalism introduced in \cite{Strelecek25}, where we demonstrated how the metric singularity caused by zero-dimensional DPs can be regularised by a~coordinate transformation into a~one-dimensional boundary. Here we demonstrate that these boundaries glue adjacent eigenstate manifolds into a~unified geometric structure. Each DP then acts similarly to an Einstein--Rosen bridge (wormhole)~\cite{Einstein35} between otherwise disconnected manifolds. This is demonstrated in Fig.~\ref{fig:illustration} around a~DP (red cross). By introducing stretched coordinates centred at each DP, we map the singular point into a~circular boundary (the \emph{DP horizon}, dashed curve in Fig.~\ref{fig:illustration}(c)) whose geometric circumference is $\mathcal{C}=\pi$. The stretched coordinates reparametrise the polar coordinates around the DP, which already regularise the metric and realise a~blow-up of the DP~\cite{Kurdyka08}, see Sec.~\ref{sec:stretched_coords}. The metric is therefore made continuous across the horizon and admits analytic continuation between the adjacent eigenstate manifolds. The resulting object---the \emph{connected state manifold} (CSM)---is the union of eigenstate manifolds glued along their DP horizons. We further classify its topology by genus and number of ends, represent it as a~graph (vertices for eigenstate manifolds, edges for DPs), and extend the local nodal-lines rules, already known from studies of conical intersections~\cite{Herzberg63,Mead92} and related to the wave dislocation lines~\cite{Berry07Topological}, to global rules on the CSM.

We illustrate the CSM framework on a~simple three-level model with multiple DPs, inspired by the effective spin Hamiltonian of nitrogen--vacancy centres in diamond~\cite{Doherty13,Rondin14}. This model is relevant for quantum state driving \cite{Kenmoe16, Kolodrubetz17, Xu18, Band22} and metrology \cite{Gonzales22, Gassab25}, where the Fisher information metric, closely related to the quantum geometric tensor, motivates our geometrical analysis \cite{Liu20}. We demonstrate three advantages of the CSM, along with its limitations. First, the metric tensor and geodesic computation in polar and stretched coordinates near DPs become numerically stable, whereas in Cartesian coordinates the metric diverges at DPs. Second, the CSM can admit \emph{geodesic shortcuts}: boundary-value geodesics that connect two points of an eigenstate manifold and are shorter than any geodesic confined to this single-state manifold. In our example model, the new geodesic uses two DPs: one to traverse into an adjacent manifold and the other to return to the manifold of origin. This new CSM geodesic is not only shorter, but offers an even stronger advantage: it is stable, whereas the single-manifold geodesic is infinitely sensitive to its initial condition. Third, we propose a~systematic approach to computing the Berry phase accumulated along adiabatic evolutions~\cite{Albash18, Ventura25} that traverses DPs in effectively~$\PT$-symmetric systems. This links the nodal line structure to the topology of the CSM and provides geometrical insights into the Berry phase.

The paper is organised as follows. First, we introduce the quantum geometric tensor in Sec.~\ref{sec:geometric_tensor}, geodesics in Sec.~\ref{sec:geodesics} and describe a~more suitable formalism for near-DP behaviour in Sec.~\ref{sec:Bloch_formalism}. We show how the metric behaves near DPs in Sec.~\ref{sec:geometric_tensor_near_DP}, specifically for 2- and $N$-level systems and discuss the difference between exact and avoided crossings. We proceed by introducing two building blocks of the CSM: the analytic continuation formalism of the wavefunction through DPs in Sec.~\ref{sec:wormholes}, and signed polar coordinates that allow for metric tensor regularisation, together with stretched coordinates that provide natural charts on the CSM. Having all pieces prepared, we formalise the connected state manifold and characterise its topological properties, including genus and number of ends in Sec.~\ref{sec:CSM}. Nodal line theory is then briefly described, along with its implications for the CSM in Sec.~\ref{sec:nodal_lines}. In Sec.~\ref{sec:applications} we apply the framework to numerical regularisation, geodesic shortcuts, and nodal-line structure. We conclude in Sec.~\ref{sec:conclusion} and discuss the limitations, possible extensions, and further applications of the CSM framework. The appendices contain technical details of the perturbation theory near DPs, the explicit calculation of the geometric tensor for a~more general 3-level system, an analytical argument for geodesic behaviour along lines with a~degenerate metric tensor and a~numerical method for nodal line computation.

\section{General theory}
\subsection{Geometric tensor}
\label{sec:geometric_tensor}
We are interested in finite quantum systems with $N$-dimensional Hilbert space described by a~class of Hermitian Hamiltonians $\H(\ccoord)$ with 2-dimensional \emph{parametric space}
\begin{equation}
    \ccoord\equiv (\coord^1, \coord^2)\in \R^2,
\end{equation}
which we take to be the whole, unbounded $\R^2$ plane (compact parametric spaces are discussed separately in Sec.~\ref{sec:compact_parametric}), and discrete spectrum $\{E_0(\ccoord), E_1(\ccoord), \ldots, E_{N-1}(\ccoord)\}$. We further assume that the spectrum remains nondegenerate except at isolated diabolic points (DPs) located at finite coordinates in the parametric space. Extended degeneracies, which require a~non-Abelian geometry formulation~\cite{Wilczek84, Ma10}, lie outside of our scope. We describe the location of DPs using $\{\ccoord_i\}_{i\in I}$ (for index set $I$) between adjacent levels, where $E_n(\ccoord_i) = E_{n+1}(\ccoord_i)$. To jointly mark the location and energy levels associated, we refer to these only using a~simple index DP$_i$. They constitute degeneracies of Hermitian Hamiltonians and are defined by linear energy splitting 
\begin{equation}
    E_{n+1}(\ccoord) - E_n(\ccoord) \propto |\ccoord-\ccoord_i|.    
    \label{eq:DP_definition_endif}
\end{equation}
The linear splitting distinguishes DPs from more general \emph{degenerate points}, where the energy difference may follow a~different power law.

For general multi-parametric Hamiltonians, DPs have codimension 3, manifesting as isolated points in 3-parametric space~\cite{Berry84Diabolical, Nesterov09}. For Hamiltonians possessing an effective $\PT$ symmetry (or \emph{antiunitary symmetry})---that is, there exists a~basis in which $\H$ is real-valued~\cite{Louvet23}---DPs codimension is 2, and they appear as isolated points in 2-parametric systems. We therefore restrict our analysis of 2-parametric systems to those possessing an effective $\PT$ symmetry.

On regions where the spectrum is nondegenerate, the parametric space can be equipped with the quantum geometric tensor $T_{\mu\nu}^{(n)}$ (QGT), induced by a~particular $n$th level eigenstate $\ket{\psi_n(\ccoord)}$ as (omitting the $\ccoord$ dependence for clarity)
\begin{equation}
    T_{\mu\nu}^{(n)}(\ccoord) \equiv \bra{\partial_\mu \psi_n}\left[\Id - \ket{\psi_n}\!\bra{\psi_n}\right]\ket{\partial_\nu \psi_n}
    = \sum_{j \neq n} \frac{\braket{\psi_n|\partial_\mu \H|\psi_j}\braket{\psi_j|\partial_\nu \H|\psi_n}}{(E_j - E_n)^2},
    \label{eq:QGT_definition}
\end{equation}
for partial derivatives $\partial_\mu$ in the $\coord^\mu$ direction. The second equality follows from inserting a~basis-expanded identity and applying the Hellmann--Feynman theorem. We primarily use the real part of this tensor, the \emph{Provost--Vallee metric tensor}: \begin{equation}
    g_{\mu\nu}^{(n)}(\ccoord) = \Re\, T_{\mu\nu}^{(n)}(\ccoord).
    \label{eq:g_definition}
\end{equation}
It defines infinitesimal distances on the $n$th state manifold $\M^n$ via $(\d s^{(n)})^2 = g_{\mu\nu}^{(n)}\d\coord^\mu \d\coord^\nu$. Throughout the article we assume the summation over repeated Greek indices. The manifold $\M^n$ is therefore a~parameter space (domain in $\R^2$) equipped with the pullback of the Provost--Vallee metric from the space of rays, where the metric is natively defined. 

The remaining imaginary part of the geometric tensor determines the \emph{Berry curvature} 
\begin{equation}
    V_{\mu\nu}^{(n)}(\ccoord) = -2\,\Im\, T_{\mu\nu}^{(n)}(\ccoord),
    \label{eq:V_definition}
\end{equation}
which is strongly influenced by the existence of the DPs.

Consider an eigenstate $\ket{\psi(0)}$ in time $t=0$, that is \emph{adiabatically} transported along a~closed path $\Gamma: [0,T]\mapsto \ccoord(t)\in\R^2$. During the time evolution driven by $\H(\ccoord(t))$, the eigenstate acquires a~phase \cite{Berry84} (assuming $\hbar=1$)
\begin{equation}
    \ket{\psi(T)} = e^{i\gamma_{\mathrm d}(T)}e^{i\gamma_{\mathrm B}(T)}\ket{\psi(0)}.
\end{equation}
This shortened notation separates the acquired phase into two parts: the \emph{dynamical phase}, defined purely by the Hamiltonian spectrum
\begin{align}
    \gamma_{\mathrm d}(T) &= -\int_0^T \braket{ \psi(t)|\H(\ccoord(t))|\psi(t)} \d t,
\end{align}
and the \emph{geometric (Berry) phase}, depending on the path $\Gamma$,
\begin{equation}
    \gamma_{\mathrm B}(T) = \oint_{\Gamma} \mathcal{A}(\ccoord) \cdot \d\ccoord = \frac{1}{2}\iint_{S} V_{\mu\nu}^{(n)}(\ccoord) \d\coord^\mu \wedge \d\coord^\nu,
    \label{eq:Berry_phase}
\end{equation}
for $\mathcal{A}_\mu (\ccoord)\equiv i\braket{\psi_n(\ccoord)|\partial_\mu\psi_n(\ccoord)}$ being the Berry connection, and $S$ being the surface bounded by $\Gamma$. The second equality comes from the Stokes' theorem and connects the Berry phase to the Berry curvature and QGT.

As $\Gamma$ traces a~path in the parametric space, the induced geometric phase can count the number of DPs enclosed: Any smooth deformation of $\Gamma$ across a~DP produces a~discontinuous jump in $\gamma_\mathrm{B}$. In systems possessing the effective $\PT$-symmetry, this jump takes a~quantised value~\cite{Herzberg63}. As a~result, the Berry phase is
\begin{equation}
    \gamma_{\mathrm B} = k\pi \mod 2\pi,
    \label{eq:Berry_phase_quantization}    
\end{equation}
where $k$ is the net number of DPs enclosed by $\Gamma$. The DPs are in this sense analogous to magnetic monopoles, providing sources of quantised geometric flux~\cite{Berry84, Louvet23}. This formula demonstrates the structural stability of DPs, in contrast to general degenerate points to which this law does not extend.

One limitation of this framework is the assumption of nondegenerate eigenstates along the entire path $\Gamma$. It does not describe what happens when $\Gamma$ passes directly through a~DP, which is the problem we solve in Sec.~\ref{sec:nodal_lines}.

\subsection{Geodesics}
\label{sec:geodesics}
Geodesics are the locally shortest paths on eigenstate manifolds. They are connected to the adiabatic quantum state driving, where the Hamiltonian parameters $\ccoord(t)$ are varied slowly along a~path in parametric space keeping the system in the instantaneous eigenstate. Practically, they can be found by solving the geodesic equation
\begin{equation}
    \frac{\d^2 \coord^\mu}{\d t^2} + \Gamma^\mu_{\;\;\nu\xi}(\ccoord) \frac{\d \coord^\nu}{\d t}\frac{\d \coord^\xi}{\d t} = 0, \quad \Gamma^\mu_{\;\;\nu\xi}(\ccoord) = \frac{1}{2}g^{\mu\lambda}(\partial_\nu g_{\lambda\xi} + \partial_\xi g_{\lambda\nu} - \partial_\lambda g_{\nu\xi}),
    \label{eq:geodesicEq}
\end{equation}
for Christoffel symbols $\Gamma^\mu_{\;\;\nu\xi}(\ccoord)$, and affine parameter $t$. This can be solved as an \emph{initial value problem} by specifying the initial point $\ccoord(t=0)$ and velocity $\ccoord'(t=0)$, or as a~\emph{boundary value problem} by specifying the initial and final points $\ccoord(t=0),\,\ccoord(t=T)$.

On a~single eigenstate manifold $\M^n$, multiple paths connecting the same two points can solve the geodesic equation, creating multiple geodesics with different lengths~\cite{Strelecek25}.

Length is measured in a~coordinate-independent way by integrating the speed $v(t)$ along the entire path
\begin{equation}
    s(t) = \int_{0}^{t} v(\tau)\d \tau , \quad v(t)=\sqrt{g_{\mu\nu}(\ccoord(t)) \frac{\d \coord^\mu(t)}{\d t}\frac{\d \coord^\nu(t)}{\d t}},
    \label{eq:geodesic_distance}
\end{equation}
for the affine parameter $t$. For geodesics, the speed is always constant.

\subsection{Bloch vector formalism}
\label{sec:Bloch_formalism}
In the geometric tensor \eqref{eq:QGT_definition}, the first expression requires derivatives of the wavefunction, which can be difficult to evaluate numerically. The second, sum-over-states expression avoids this by shifting the derivatives to the Hamiltonian, but the bottleneck of computing the eigenstates remains. To circumvent this obstacle, we employ the Bloch-vector formalism developed in Refs.~\cite{Pozo20, Graf21}, which expresses the quantum geometric tensor entirely in terms of the Hamiltonian and its eigenvalues, and is often simpler than the eigenvector-based approach. 

The formalism relies on a~generalised Bloch-vector representation of quantum states and expanding the Hamiltonian in the generators $\{\hat\lambda_k\}_{k=1}^{N^2-1}$ of the Lie algebra associated with the group $SU(N)$, which can be assembled with some fixed ordering into a~vector $\bm{\hat\lambda}$. A~general $N$-dimensional Hermitian Hamiltonian can be expressed as
\begin{equation}
    \H(\ccoord) = h^0 \Id+\sum_{k=1}^{N^2-1} h^k(\ccoord)\,\hat\lambda_k \equiv h^0 \Id+\bm h(\ccoord)\cdot \bm{\hat\lambda},
    \label{eq:H_generator_expansion}
\end{equation}
where the energy offset $h^0\Id$ does not influence eigenstates, keeping the geometry intact, and can be omitted without the loss of generality. Without this term, the Hamiltonian can be written more compactly in a~traceless form as a~\emph{Hamiltonian vector}
\begin{equation}
    \h(\ccoord) = (h^1(\ccoord), \ldots, h^{N^2-1}(\ccoord))^T, \quad h^k(\ccoord) = \frac{1}{2}\Tr(\H(\ccoord) \hat\lambda_k).
    \label{eq:Hamiltonian_vector}
\end{equation}
The $n$th-level eigenprojector is then directly related to one of the possible definitions of the Bloch vector of the $n$th eigenstate $\bm b^{(n)}(\ccoord)$, see \cite{Graf21} for details,
\begin{equation}
    \hat P^{(n)}(\ccoord) \equiv\ket{\psi_n(\ccoord)}\!\bra{\psi_n(\ccoord)} = \frac{1}{N}\Id + \frac{1}{2}\bm{b}^{(n)}(\ccoord) \cdot \hat{\bm{\lambda}} \quad \Rightarrow \quad \bm b^{(n)}(\ccoord) = \Tr(\hat P^{(n)}(\ccoord)\,\hat{\bm{\lambda}}).
    \label{eq:eigenprojector_expansion}
\end{equation}
Finally, the geometric tensor is
\begin{equation}
    T_{\mu\nu}^{(n)}(\ccoord) = \underbrace{\frac{1}{4}\partial_\mu \bm b^{(n)}(\ccoord) \cdot \partial_\nu \bm b^{(n)}(\ccoord)}_{g_{\mu\nu}^{(n)}(\ccoord)} + \frac{i}{4}\underbrace{\bm b^{(n)}(\ccoord) \cdot \left(\partial_\mu \bm b^{(n)}(\ccoord) \times_L \partial_\nu \bm b^{(n)}(\ccoord)\right)}_{-2V_{\mu\nu}^{(n)}(\ccoord)},
    \label{eq:geometric_tensor_Bloch}
\end{equation}
using the generalised cross products in Lie algebra described in Eq.~\eqref{eq:structure_constants}.

Crucially, the Bloch vector corresponding to the $n$th eigenstate, $\bm b^{(n)}(\ccoord)$, can be constructed directly from the Hamiltonian vector $\h(\ccoord)$, its derivatives, and the eigenvalue $E_n(\ccoord)$, without requiring the explicit computation of eigenvectors. We demonstrate this for two- and $N$-level systems and show what it means for the geometry near the DPs.

\subsection{Quantum geometric tensor near diabolic points}
\label{sec:geometric_tensor_near_DP}
Suggested by the summation form in Eq.~\eqref{eq:QGT_definition}, DPs generally cause divergences in the metric tensor elements as $E_{n} \to E_{n+1}$. These can be regularised by a~careful choice of coordinates \cite{Strelecek25}.

Here we study the general behaviour of the metric tensor near the DP. We transform the Hamiltonian in Eq.~\eqref{eq:H_generator_expansion} from its native coordinates $\ccoord$ into \emph{polar coordinates} around some $\mathrm{DP}_i$ chosen by the index $i$. Polar coordinates, defined by $\ccoord-\ccoord_i=r (\cos\phi,\sin\phi)$, respect the system's symmetry and substantially simplify the metric tensor, because the $\mathrm{DP}_i$ can be approached by a~single parameter change $r\rightarrow 0$. Without loss of generality, we choose a~DP between $\M^0$ and $\M^1$, which implies the following behaviour of the spectrum:
\begin{equation}
    E_k(r,\phi)-E_0(r,\phi) = \begin{cases}
        r\Delta_1(\phi) +\O(r^2), & k=1 \\
        \Delta_k + \O(r), & k=2, \ldots, N-1,
    \end{cases}
    \label{eq:spectrum_linearization}
\end{equation}
for energy offsets $\Delta_k>0$ and $\Delta_1(\phi) > 0$ for all $\phi$, as required by Eq.~\eqref{eq:DP_definition_endif}. In these coordinates, the Hamiltonian can be expanded as
\begin{equation}
    \bm h(r,\phi) = \bm{\tilde f} + r \bm f^{\{1\}}(\phi) + r^2 \bm f^{\{2\}}(\phi) + \O(r^3),
    \label{eq:H_expansion_DP}
\end{equation}
where $\bm{\tilde f}\equiv\lim_{r\to 0}\bm h(r,\phi)$ determines the spectrum at the DP and term $\bm f(\phi)\equiv \bm f^{\{1\}}(\phi)$ is a~vector of functions capturing the leading angular dependence. We require the coefficient functions $\bm f^{\{1\}}(\phi)$ and $\bm f^{\{2\}}(\phi)$ to be real and continuously differentiable in $\phi\in[0,2\pi)$. 

The Hamiltonian expansion \eqref{eq:H_expansion_DP} up to the second order expresses the geometric tensor in the limit $r\to 0$, as shown in Appendix~\ref{sec:appendix_SPT}.
 
\subsubsection{Two-level systems}
\label{sec:two_level}
For $N=2$ system the canonical choice for generators are Pauli matrices $\hat{\bm\lambda} = (\hat\sigma_1, \hat\sigma_2, \hat\sigma_3)^T$, in which the general Hamiltonian \eqref{eq:Hamiltonian_vector} is expressed through
\begin{equation}
    \h(\ccoord) = (h^1(\ccoord), h^2(\ccoord), h^3(\ccoord))^T.
    \label{eq:2level_hamiltonian_vector}
\end{equation}
Its eigenvalues are $E_1 = |\h|= -E_0 $, with respective Bloch vectors $\bm b^{1} = \h/|\h|= -\bm b^{0} $~\cite{Graf21}.

Without loss of generality, we restrict to the 2D plane, for which the Hamiltonian is real-valued \cite{Matus23}. The remaining two coordinates can then be \emph{aligned with the Hamiltonian parameters}, leading to
\begin{equation}
    h^1(\phi)\equiv x= r\cos\phi, \qquad h^2(\phi) = 0,\qquad  h^3(\phi)\equiv z=r\sin\phi.
    \label{eq:two-level-alignment}
\end{equation}
This sets the DP to $r = 0$. The metric tensor from Eq.~\eqref{eq:geometric_tensor_Bloch} in Cartesian coordinates reads
\begin{equation}
    g^{(n)}_{\mu\nu}(x,z) \equiv \begin{pmatrix}
        g_{xx} & g_{xz} \\
        g_{zx} & g_{zz}
    \end{pmatrix}= \frac{1}{4(x^2+z^2)^2}
    \begin{pmatrix}
        z^2 & -xz \\
        -xz & x^2
    \end{pmatrix},
    \label{eq:2level_metric_cartesian}
\end{equation}
which exhibits divergences as $(x,z) \to (0,0)$. In polar coordinates, the degenerate behaviour of the whole metric reveals itself, leading to a~metric tensor with a~well-defined limit to the DP:
\begin{equation}
    g^{(n)}_{\mu\nu}(r,\phi) \equiv \begin{pmatrix}
        g_{rr} & g_{r\phi} \\
        g_{\phi r} & g_{\phi\phi}
    \end{pmatrix} = 
    \begin{pmatrix}
        0 & 0 \\
        0 & \tfrac{1}{4}
    \end{pmatrix}.
    \label{eq:2level_metric_polar}
\end{equation}
The divergence of metric components is therefore a~coordinate artefact, removable by passing to polar coordinates. The vanishing of $g_{rr}$ reflects that the Bloch vector depends only on the direction of the Hamiltonian vector, not its magnitude, leading to $\partial_r \bm b=0$. This is equivalent to the eigenvector being independent of the radial coordinate $r$, since at an exact level crossing the two levels do not interact.

To clarify where the divergences in the Cartesian metric tensor in the vicinity of DPs originate, we briefly discuss the case of avoided crossings. We do this in a~single-parametric 2-level system~\eqref{eq:2level_hamiltonian_vector}, interpreting the second variable as an energy gap $z\equiv \Delta = (E_1-E_0)/2$ at $x=0$. The element $g_{xx}^{(n)} = \Delta^2/[4(x^2 + \Delta^2)^2]$ then exhibits qualitatively different limits depending on the order:
\begin{equation}
    \lim_{\Delta \rightarrow 0} \lim_{x \rightarrow 0} g_{xx}^{(n)} = \infty, \quad \lim_{x \rightarrow 0} \lim_{\Delta \rightarrow 0} g_{xx}^{(n)} = 0.
    \label{eq:limits_metric_2level}
\end{equation}
The first limit describes an avoided crossing: the eigenstate changes rapidly with $x$, and the sensitivity diverges as the gap closes. The second describes an exact diabolic crossing, where the coupling between the two levels vanishes and the metric with it.

\subsubsection{Multi-level systems}
\label{sec:n_level}
For $N > 2$ systems with Hamiltonian expansion \eqref{eq:H_expansion_DP}, the metric tensor \eqref{eq:QGT_definition} receives contributions from levels that do not participate in the degeneracy, here chosen between $\M^n$ and $\M^{n+1}$. These contributions generically prevent $g_{rr}$ from vanishing identically, which is qualitatively different from the $N=2$ case. To derive explicit metric tensor elements in the DP vicinity, we use perturbation theory, see Appendix~\ref{sec:appendix_SPT}. For clarity, we also made an explicit analytical calculation for the 3-level case using the Bloch vector formalism, see Appendix~\ref{sec:appendix_3level_details}. 

The geometric tensor in the DP vicinity exhibits a~hierarchy of level contributions in the limit $r\to 0$ (see Eqs.~\eqref{eq:Tphiphi_result},~\eqref{eq:Trr_result}, and~\eqref{eq:Trphi_result}):
\begin{itemize}
    \item $\tilde T_{\phi\phi}^{(n)}$: only the degenerate pair (term $j=1$ from Eq.~\eqref{eq:QGT_definition}) contributes. It can be determined from the first-order Hamiltonian expansion \eqref{eq:H_expansion_DP}.
    \item $\tilde T_{r\phi}^{(n)}$: only the degenerate pair contributes, but it requires a~second-order Hamiltonian expansion to be evaluated, which implicitly involves coupling to all levels.
    \item $\tilde T_{rr}^{(n)}$: all eigenlevels contribute; both the degenerate pair (via second-order Hamiltonian term) and distant levels directly.
\end{itemize} 
For the angular metric element, contributions from levels not participating in the degeneracy are suppressed by $\O(r^2)$, so that $\tilde{g}_{\phi\phi}^{(n)}$ is determined solely by the two-level subsystem at the DP. To make this explicit, we order the $SU(N)$ generators such that the first three, $(\hat\lambda_1, \hat\lambda_2, \hat\lambda_3)$, form the $SU(2)$ subalgebra of the degenerate subspace $\{\ket{\psi_n}, \ket{\psi_{n+1}}\}$ at the DP. The Hamiltonian \eqref{eq:Hamiltonian_vector} is then decomposed as
\begin{equation}
    \bm h(\ccoord) = (\underbrace{h^1, h^2, h^3}_{\bm h_\times}, h^4, \ldots, h^{N^2-1})^T,
    \label{eq:hsubvector}
\end{equation}
where the cross marks the energy crossing. 

The linear order from \eqref{eq:H_expansion_DP} of the 2-level subspace, marked consistently as $\tilde{\bm f}_\times+\bm f_\times^{\{1\}}(\phi)$, then fully determines the angular metric tensor component
\begin{equation}
    \tilde g_{\phi\phi}^{(n)}(\phi) \equiv\lim_{r\rightarrow 0}g_{\phi\phi}^{(n)}(r,\phi) = \frac{1}{4}\left\|\frac{\hhsub^{\{1\}}(\phi)}{|\hhsub^{\{1\}}(\phi)|} \times \partial_\phi\left(\frac{\hhsub^{\{1\}}(\phi)}{|\hhsub^{\{1\}}(\phi)|}\right)\right\|^2,
    \label{eq:limitgphiphi}
\end{equation}
for the Cartesian cross product in $\R^3$. This can be further simplified by \emph{aligning} the coordinate system with the generators as defined in Eq.~\eqref{eq:two-level-alignment}, leading to the same result as the 2-level system
\begin{equation}
    \tilde g_{\phi\phi}^{(n)} = 1/4.
    \label{eq:limitgphiphi_aligned}
\end{equation}
For an effectively $\PT$-symmetric $\H$, such coordinates always exist. The two-level block near the DP in linear order is $h^1\hat\sigma_1+h^3\hat\sigma_3$, where $h^1$ and $h^3$ are linear functions of $\ccoord-\ccoord_i$. Choosing the new coordinates $(x,z)=(h^1,h^3)$ gives the aligned form~\eqref{eq:two-level-alignment}, up to higher orders that do not enter Eq.~\eqref{eq:limitgphiphi}. These are valid coordinates, because linear energy splitting in every direction means that $h^1$ and $h^3$ vanish together only at the DP. Eqs.~\eqref{eq:limitgphiphi} and~\eqref{eq:limitgphiphi_aligned} were derived for the linearised 3-level system, see Appendix~\ref{sec:appendix_3level_details}. However, because the angular metric depends only on the two-level subsystem and on the first-order term in the Hamiltonian expansion \eqref{eq:H_expansion_DP}, they are valid even for the $N$-level system with nonlinear terms.

The remaining metric components also have finite limits at the DP, see Appendix~\ref{sec:appendix_metric_components}. In aligned polar coordinates around the DP, the metric takes the form (cf. Eq.~\eqref{eq:2level_metric_polar})
\begin{equation}
    g^{(n)}_{\mu\nu}(r,\phi) =
    \begin{pmatrix}
        \tilde g_{rr}^{(n)}(\phi) + \mathcal O(r)& \tilde g_{r\phi}^{(n)}(\phi) + \mathcal O(r) \\
        \tilde g_{r\phi}^{(n)}(\phi) + \mathcal O(r) & \tfrac{1}{4}+ \mathcal O(r)
    \end{pmatrix},
    \label{eq:Nlevel_metric_polar}
\end{equation}
for $\tilde g_{rr}^{(n)}(\phi)\geq 0$ and $\tilde g_{r\phi}^{(n)}(\phi)$ given for the ground state manifold as real parts of Eqs.~\eqref{eq:Trr_result},\eqref{eq:Trphi_result}. The off-diagonal limit $\tilde g_{r\phi}^{(n)}(\phi)$ is generally nonzero, see Eq.~\eqref{eq:limitgphiphiRealAligned}, and can be removed by a~local diagonalisation of the metric, see Appendix~\ref{sec:appendix_metric_diagonalization}.

\begin{figure}
    \centering
    \includegraphics{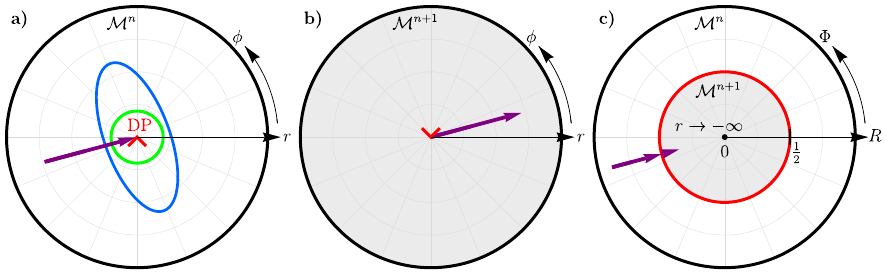}
    \caption{Schematic illustration of coordinate systems around a~DP. (a)~Manifold $\M^n$ in polar coordinates $(r,\phi)$: the DP is marked by the lower half of a~red cross. The green circle represents an $r=\mathrm{const.}$ loop in aligned coordinates, along which $g_{\phi\phi}\rightarrow 1/4$ as $r\rightarrow 0$. The blue ellipse shows a~loop of constant radius in misaligned linear coordinates, along which $g_{\phi\phi}$ varies. The lengths of both loops converge to $\CC=\pi$ as they shrink to the DP. The purple arrow depicts an example path approaching the DP. (b)~Manifold $\M^{n+1}$ (emphasised by the grey background) in the same coordinates: the path continues out of the DP, marked by the upper half of the cross; together, the two halves in~(a) and~(b) form the full DP crossing symbol. (c)~Stretched coordinates $(R,\Phi)$: the DP is expanded into a~circular boundary of radius $\tfrac{1}{2}$ (red circle), with $\M^{n+1}$ inside (grey) and $\M^n$ outside; the purple line shows the same path crossing the boundary, rescaled according to Eq.~\eqref{eq:shifted_polar_coords}.}
    \label{fig:illustration_coordinates}
\end{figure}

From the angular metric tensor element, we further define the \emph{geometric circumference of the DP}
\begin{equation}
    \CC \equiv \lim_{r\rightarrow 0} \int_0^{2\pi} \sqrt{g_{\phi\phi}}\,\d \phi = \frac{1}{2}\int_0^{2\pi}||\partial_\phi \bm b^{(0)}||\,\d\phi.
    \label{eq:circumference_definition}
\end{equation}
What constitutes a~circle in parametric space depends on the chosen coordinates, as schematically shown in Fig.~\ref{fig:illustration_coordinates}(a). The circumference of the DP from Eq.~\eqref{eq:limitgphiphi} depends solely on the curve traced by the normalised vector $\hhsub/||\hhsub||$ in 3D space as $\phi$ varies from $0$ to $2\pi$. Specifically, it equals half the length of this curve, i.e.\ the area of the cone connecting it to the origin. If $\H$ is continuously differentiable around the DP in the chosen coordinates, then $\hhsub^{\{1\}}(\phi)$ is a~linear combination of $\cos\phi$ and $\sin\phi$ with linearly independent coefficient vectors (otherwise the levels would not split linearly in some direction). The normalised vector $\hhsub/||\hhsub||$ then runs once around a~great circle, with its angle changing monotonically in $\phi$. The circumference of the DP in any such coordinates is therefore
\begin{equation}
    \CC = \pi.
    \label{eq:c_is_pi}
\end{equation}
This is formally shown in Appendix~\ref{sec:appendix_circumference} by inserting Eq.~\eqref{eq:limitgphiphi} into Eq.~\eqref{eq:circumference_definition} and performing the integration.

In Eq.~\eqref{eq:2level_metric_polar}, we have shown that for a~two-level system the radial metric tensor element vanishes, as opposed to the $N$-level system. This clarifies the limitations of the two-level approximations to the diabolic crossing, as shown for example in Ref.~\cite{Strelecek25}. The general idea is to find an effective 2-level Hamiltonian $\H_{\rm eff}(x,z)$ that can faithfully approximate the metric tensor in the vicinity of the DP for a~larger $N$-level system. This can be done in Cartesian coordinates, where the avoided-crossing regime dominates; however, in polar coordinates, the approximation can capture only the angular component $g_{\phi\phi}$. This is because for $N>2$, the components $g_{rr}$ and $g_{r\phi}$ receive nonzero contributions from distant levels through second-order eigenvector corrections (Eqs.~\eqref{eq:Trr_result},~\eqref{eq:Trphi_result})---corrections that are absent by construction in a~pure 2-level model, and hence in the $\H_{\rm eff}(x,z)$ as well.

\subsection{Analytic continuation through diabolic points}
\label{sec:wormholes}
Energy crossings at DPs introduce discontinuities to the metric tensor of individual state manifolds. This is because the metric depends on the $n$th eigenstate and its derivatives, which switch order at the DP---the $n$th eigenstate becomes $(n+1)$st and vice versa. We show how the $\M^n$ manifold can be smoothly joined with $\M^{n+1}$ by using the DP as a~bridge connecting them. In 2-parametric spaces, this is conceptually similar to wormholes in general relativity. This is a~crucial step in regularising the metric tensor around the DP and requires careful treatment.

\subsubsection{Single-parametric systems}
\label{sec:1D_continuation}
\begin{figure}
    \centering
    \begin{tikzpicture}[scale=1.0, >=stealth]
        \def\yaxis{-1.2}      
        \def\yupper{2.7}      
        \def\ylower{-0.6}     
        \def\ycross{1.0}      
        \def\slope{0.25}      
        \def\xhalf{3.5}       
        
        \pgfmathsetmacro{\yBL}{\ycross - \slope*\xhalf}   
        \pgfmathsetmacro{\yTL}{\ycross + \slope*\xhalf}   
        
        \draw[->, thick] ({-\xhalf-0.5}, \yaxis) -- ({\xhalf+0.8}, \yaxis)
            node[right] {$x$};
        \draw[thick] (0, {\yaxis+0.1}) -- (0, {\yaxis-0.1});
        \node[below] at (0, {\yaxis-0.03}) {$0$};
        
        \draw[gray, thick] (-\xhalf, \yupper) to[out=-5, in=185]
            (\xhalf, \yupper) node[right, gray] {$\M^{n+2}$};
        \draw[gray, thick] (-\xhalf, \ylower) to[out=5, in=175]
            (\xhalf, \ylower) node[right, gray] {$\M^{n-1}$};
        \pgfmathsetmacro{\ydotsup}{\yupper + 0.3}
        \pgfmathsetmacro{\ydotsdn}{\ylower - 0.13}
        \node[gray] at (0, \ydotsup) {$\vdots$};
        \node[gray] at (0, \ydotsdn) {$\vdots$};
        
        \draw[red, very thick]  (-\xhalf, \yBL) -- (0, \ycross);
        \draw[MyBlue, very thick] (0, \ycross) -- (\xhalf, \yTL)
            node[right, MyBlue] {$\M^{n+1}$};
        \draw[MyBlue, very thick] (-\xhalf, \yTL) -- (0, \ycross);
        \draw[red, very thick]  (0, \ycross) -- (\xhalf, \yBL)
            node[right, red] {$\M^{n}$};
        
        \fill (0, \ycross) circle (2.5pt);
        
        \pgfmathsetmacro{\xstart}{2.0}  
        \pgfmathsetmacro{\xend}{0.2}    
        %
        \pgfmathsetmacro{\yAs}{\ycross - \slope*\xstart-0.1}
        \pgfmathsetmacro{\yAe}{\ycross - \slope*\xend-0.1}
        \draw[->, thick] (-\xstart, \yAs) -- (-\xend, \yAe);
        \pgfmathsetmacro{\yAmid}{(\yAs+\yAe)/2}
        \node[below] at ({-(\xstart+\xend)/2-0.2}, \yAmid-0.2)
            {$\displaystyle\lim_{x\to 0^-}\!\ket{\psi_{n}(x)}$};
        \pgfmathsetmacro{\yBs}{\ycross + \slope*\xstart+0.1}
        \pgfmathsetmacro{\yBe}{\ycross + \slope*\xend+0.1}
        \draw[->, thick] (\xstart, \yBs) -- (\xend, \yBe);
        \pgfmathsetmacro{\yBmid}{(\yBs+\yBe)/2}
        \node[above] at ({(\xstart+\xend)/2+0.2}, \yBmid+0.2)
            {$\displaystyle\lim_{x\to 0^+}\!\ket{\psi_{n+1}(x)}$};
        
        \pgfmathsetmacro{\yCs}{\ycross + \slope*\xstart +0.1}
        \pgfmathsetmacro{\yCe}{\ycross + \slope*\xend +0.1}
        \draw[->, thick] (-\xstart, \yCs) -- (-\xend, \yCe);
        \pgfmathsetmacro{\yCmid}{(\yCs+\yCe)/2}
        \node[above] at ({-(\xstart+\xend)/2-0.2}, \yCmid+0.2)
            {$\displaystyle\lim_{x\to 0^-}\!\ket{\psi_{n+1}(x)}$};
        \pgfmathsetmacro{\yDs}{\ycross - \slope*\xstart -0.1}
        \pgfmathsetmacro{\yDe}{\ycross - \slope*\xend -0.1}
        \draw[->, thick] (\xstart, \yDs) -- (\xend, \yDe);
        \pgfmathsetmacro{\yDmid}{(\yDs+\yDe)/2}
        \node[below] at ({(\xstart+\xend)/2+0.2}, \yDmid-0.2)
            {$\displaystyle\lim_{x\to 0^+}\!\ket{\psi_{n}(x)}$};
    \end{tikzpicture}
    \caption{Illustration of a~diabolic level crossing in a~single-parametric system. The eigenstate manifolds $\M^n$ (red) and $\M^{n+1}$ (blue) cross at the diabolic point (DP) at $x=0$, while the distant levels $\M^{n-1}, \M^{n+ 2},\ldots$ (grey) remain separated. The arrows indicate the eigenstate limits as $x\rightarrow 0$, as given in Eq.~\eqref{eq:limitsToDP1D}: the wavefunctions exchange between the two manifolds at the crossing.}
    \label{fig:illustration_crossing}
\end{figure}
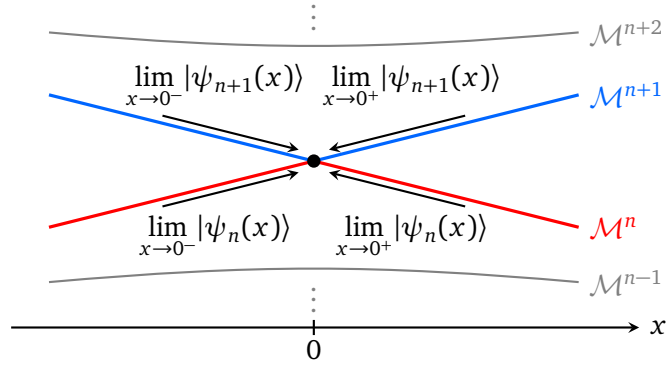

For simplicity, we start with a~single-parametric Hamiltonian $\H(x)$ for $\ccoord\equiv x\in\R$ with DP at $x=0$. Around such DP, the $U(1)$ gauge of the wavefunction can be locally fixed such that (see Fig.~\ref{fig:illustration_crossing})
\begin{equation}
    \lim_{x \to 0^-} \ket{\psi_n(x)} = \lim_{x \to 0^+} \ket{\psi_{n+1}(x)}, \quad
    \lim_{x \to 0^-} \ket{\psi_{n+1}(x)} = \lim_{x \to 0^+} \ket{\psi_n(x)}.
    \label{eq:limitsToDP1D}
\end{equation}

This formalism avoids the issue of defining the metric at the DP, as it relies only on limits approaching the DP from either side. In a~genuine diabolic crossing the coupling between the two levels vanishes, so the levels cross exactly and the eigenvectors satisfy Eq.~\eqref{eq:limitsToDP1D}. For a~Hamiltonian analytic in the real parameter $x$, this is guaranteed by Rellich's theorem~\cite{Rellich69,Kato95}: an orthonormal set of eigenvectors can be chosen analytic in $x$ even at the energy crossing. Consequently, following the $n$th eigenstate across $x=0$ leads to what is labelled the $(n+1)$st eigenstate on the other side: the manifolds $\M^n$ and $\M^{n+1}$ exchange, and each manifold is discontinuous when viewed in isolation. This allows us to construct a~continuous manifold by taking the union of two submanifolds defined on either side of the DP
\begin{equation}
    \M^{(n,n+1)}\equiv\bigl(\overline{\M^n|_{x<0}} \cup \overline{\M^{n+1}|_{x>0}}\bigr)/\sim,
    \label{eq:connected_manifold_1D}
\end{equation}
and analogously $\M^{(n+1,n)}$ with the eigenlevels switched. Here, the geometry at $x=0$ is defined by the closure (denoted by an overline), taken via the appropriate one-sided limits. The equivalence relation $\sim$ here only prevents double-counting of points by identifying the left and right limits at $x=0$ as the same point on the manifold, according to Eq.~\eqref{eq:limitsToDP1D}. The metric tensor on the manifolds constructed this way is smooth at the DP, enabling continuous geometric paths to traverse the degeneracy.

\subsubsection{Two-parametric systems and stretched coordinates}
\label{sec:stretched_coords}
In the 2-parametric system $\H(r,\phi)$ with the DP at $r=0$, metric tensor in any direction $\phi=\text{const.}$ behaves as a~single-parametric system. However, the angular coordinate $\phi$ provides the crucial new feature: a~continuous path can connect any two points on $\M^n$ by traversing around the DP.

In polar coordinates, the metric around the DP depends nontrivially on the angle $\phi$:
\begin{equation}
    \lim_{r \to 0} g_{\mu\nu}^{(n)}(r,\phi) = \tilde{g}_{\mu\nu}^{(n)}(\phi),
    \label{eq:two_param_limit}
\end{equation}
with \emph{tilde} marking the limit. This angular dependence reflects the non-uniqueness of eigenvector extensions to $r=0$, rendering the metric ill-defined at the DP itself. Rellich already showed this obstruction~\cite{Rellich69}, see also~\cite{Kato95}, when the eigenvalues $\pm\sqrt{x^2+z^2}$ of the aligned Hamiltonian~\eqref{eq:two-level-alignment} are not differentiable at the DP. Along each line  through the DP, the two levels simply cross, with eigenvalues $\pm r$ and eigenvectors independent of $r$. The limit of the ground state as $r$ approaches the DP, however, depends on the direction of approach $\phi$, so the ground state has no continuous extension to the DP.

To circumvent this problem, we seek local coordinates that (i) include the DP itself into the chart while retaining a~well-defined limiting geometry, and (ii) make the DP act as a~boundary between the two adjacent eigenstate manifolds. Such a~boundary allows us to connect the two manifolds through the DP, which leads to the global structure of the CSM in Sec.~\ref{sec:CSM}.

Polar coordinates $(r,\phi)$ centred at the DP already achieve (i). In them, the DP is not a~single point but the set $r=0$ of all directions $\phi$, on which the limits~\eqref{eq:two_param_limit} are well defined. To achieve (ii), we extend the radial coordinate to a~signed variable $r\in\R$. Its absolute value is the distance from the DP, $\ccoord-\ccoord_i=|r|(\cos\phi,\sin\phi)$, and its sign tells us whether we are on $\M^n$ ($r>0$) or on $\M^{n+1}$ ($r<0$). A~straight path through the DP continues from the direction $\phi$ on $\M^n$ to the opposite direction $\phi+\pi$ on $\M^{n+1}$, see the purple arrow in Fig.~\ref{fig:illustration_coordinates}. We therefore measure the angle on $\M^{n+1}$ from the opposite direction, $\Phi=\phi+\pi$, while $\Phi=\phi$ on $\M^n$, so that the straight path through the DP has constant coordinate $\Phi$ while $r$ passes through zero. We call $(r,\Phi)$ the \emph{signed polar coordinates}. Without the signed $r$, the off-diagonal metric tensor element would acquire the opposite sign on $\M^{n+1}$ when expressed in the same coordinate orientation. Since this chart transforms the DP into the line $r=0$, it removes the singularity and regularises the metric at the DP.

Already the polar coordinates are a~blow-up of the DP, replacing it by the circle of directions from which it is approached. Blow-ups make the eigenvalues and an orthonormal set of eigenvectors of real-analytic Hamiltonians analytic locally~\cite{Kurdyka08}, and at a~DP, due to the linear splitting~\eqref{eq:DP_definition_endif}, this single blow-up suffices. The eigenvectors cannot be chosen analytic globally, since following an eigenvector once around the DP changes its sign. In the signed polar coordinates, the continuation through the DP is satisfied as the eigenstate is continued along each line $\Phi=\mathrm{const.}$, $r\in\R$, as discussed above.

To obtain a~more faithful picture, we introduce what we call \emph{stretched coordinates} $(R,\Phi)$. They compactify $\M^{n+1}$ into a~disc and draw the DP as a~circle of the correct length, which is an intermediate step towards picturing the DP as a~bridge between the two manifolds, as sketched in Fig.~\ref{fig:illustration}(c). They are defined by
\begin{equation}
    r(R) = R-\frac{1}{4R}, \quad
    \phi(R,\Phi) = \begin{cases}
        \Phi & R > \frac{1}{2}, \\
        \Phi + \pi & R < \frac{1}{2}.
    \end{cases}
    \label{eq:shifted_polar_coords}
\end{equation}
This transformation maps the DP from $r=0$ to the circle $R = \frac{1}{2}$, which we term the \emph{DP horizon}. The exterior region $R > \frac{1}{2}$ corresponds to $\M^n$ near the DP, while $\M^{n+1}$ is mapped to the interior $R < \frac{1}{2}$ with inverted orientation, see Fig.~\ref{fig:illustration_coordinates}(c), and its limit $r \to -\infty$ is projected to the centre $R = 0$. The $\pi$ shift in $\phi$ accounts for the coordinate inversion in two dimensions, and the DP horizon radius $\frac{1}{2}$ follows from Eq.~\eqref{eq:c_is_pi}, ensuring the DP horizon length matches the geometric DP circumference. Since $r(R)$ is increasing, this transformation is only a~reparametrisation of the radial coordinate, which leaves the smoothness of the metric unchanged.

Because a~well-defined limit towards the DP in signed polar coordinates ensures smooth continuation of the wavefunction across the DP horizon, the signed radial coordinate guarantees continuity of radial derivatives up to the gauge choice
\begin{equation}
    \lim_{r\rightarrow 0^+}\partial_{r} \ket{\psi^{(n)}(r,\Phi)} = \lim_{r\rightarrow 0^-}\partial_{r} \ket{\psi^{(n+1)}(r,\Phi)}.
    \label{eq:radial_psi_derivative_continuity}
\end{equation}
We can assume the Hamiltonian is $p$-times continuously differentiable $\hat H(\ccoord)\in C^p$ for some $p \geq 2$. The projector onto both crossing levels (Kato's total projection, also called the Riesz projector) is then $C^p$ through the DP, unlike the eigenprojectors of the individual levels~\cite{Kato95}. Within this two-level subspace, the continued state is an eigenvector of the traceless part of the Hamiltonian divided by the signed $r$, which, unlike the Hamiltonian itself, remains non-degenerate at the DP. The division by $r$ costs one derivative. Hence the eigenprojector of the continued state, $\hat P^{(n)}(r,\Phi)$ for $r>0$ and $\hat P^{(n+1)}(r,\Phi)$ for $r<0$, as well as $\hat P^{(n)}(R,\Phi)$, are of class $C^{p-1}$ on a~neighbourhood of the DP horizon, and the metric $g_{\mu\nu}(R,\Phi)$, being expressible in terms of its first derivatives, belongs to $C^{p-2}$ on that domain. For an analytic $\H$, as in all examples of this paper, the projector and the metric are analytic.

The \emph{extended metric tensor} in $(R,\Phi)$ coordinates on $\M^{(n,n+1)}$ from Eq.~\eqref{eq:connected_manifold_1D} is then
\begin{equation}
    g_{\mu\nu}^{(n,n+1)}\equiv \begin{cases}
        g_{\mu\nu}^{(n)} & R>1/2\\
        \lim\limits_{R \to 1/2^+} g_{\mu\nu}^{(n)} = \lim\limits_{R \to 1/2^-} g_{\mu\nu}^{(n+1)} & R=1/2\\
        g_{\mu\nu}^{(n+1)} & R<1/2
    \end{cases}
    \label{eq:extended_metric}
\end{equation}

As opposed to the single-parametric systems, the introduction of a~second parameter allows connecting the two sides of a~DP (the directions $\phi$ and $\phi+\pi$) on any individual manifold $\M^n$. The construction provides smooth local coordinates around DPs, opening new ways for studying the geometry in the DP vicinity.

\subsection{Connected state manifold}
\label{sec:CSM}
Having established analytic continuation through DPs, we now introduce the \emph{connected state manifold} (CSM) as the union of individual eigenstate manifolds glued along their DP horizons. Intuitively, one may drive the system adiabatically on a~given eigenstate manifold and, upon reaching a~DP horizon, continue smoothly onto the adjacent manifold, so that these sheets form a~single connected geometric object. Here the term \emph{adiabatic driving} is loosened to allow for exact level crossings, where the state $\ket{\psi(t)}$ remains an exact eigenstate during the whole driving protocol. For a~rigorous treatment, see Ref.~\cite{Zhu22}. 

\subsubsection{CSM definition}
\label{sec:CSM_definition}
For $\H(\ccoord)$ introduced in Eq.~\eqref{eq:H_generator_expansion} we denote the set of all DP coordinates
\begin{equation}
    \mathcal D = \{\ccoord_\dpind\}_{\dpind \in I}, \quad I=\{\underbrace{q_0,\dots, q_a}_{I_0},\underbrace{q_b\dots q_c}_{I_1},\dots\},
    \label{eq:DP_set}
\end{equation}
where the index set $I$ naturally decomposes to individual eigen-levels $I=\{I_n\}_{n=0}^{N-1}$, where $E_n(\ccoord_\dpind) = E_{n+1}(\ccoord_\dpind)$. $I_0$ therefore holds indices of all DPs between $\M^0$ and $\M^1$, $I_1$ between $\M^1$ and $\M^2$ etc. We further require: 
\begin{enumerate}
    \item There are no DPs between non-adjacent levels. This excludes the case where three eigenlevels cross at the same point.
    \item Each degenerate point must be a~proper DP according to Eq.~\eqref{eq:DP_definition_endif}.
\end{enumerate}
Loosening the first condition is possible in some cases\footnote{The degenerate point can bridge $\M^n$ and $\M^{n+2}$ directly, leaving $\M^{n+1}$ unaffected, provided that the middle level does not interact with the crossing, i.e.\ decouples from the other two levels at first order in $\ccoord-\ccoord_\dpind$. An example of such a~triple crossing is $\H=(x\hat\sigma_1+z\hat\sigma_3)\oplus\lambda(x,z)$ with $\lambda(0,0)=0$ and $|\lambda(x,z)|<\sqrt{x^2+z^2}$ elsewhere near the origin, for which the point is an ordinary DP between $\M^n$ and $\M^{n+2}$ with $\CC=\pi$. However, the middle level can interact with the crossing, making it a~proper triple crossing, which in general does not create a~DP horizon of circumference $\CC=\pi$, as in the rotating shallow-water model~\cite{Ganeshan26}.}, in which CSM with different topology can be constructed. Loosening the second condition is more problematic: if the levels do not split linearly in every direction, as in Eq.~\eqref{eq:spectrum_linearization}, the eigenstate reached at the DP can change abruptly with the direction of approach, so that the metric need not extend continuously to the DP horizon. In directions in which the two levels touch instead of crossing, the eigenvector stays on the same manifold. This deforms the DP bridges and possibly closes them entirely in a~sense that no path can cross between eigenstate manifolds.  

Around each DP in $\mathcal D$ we locally introduce stretched coordinates $(R_\dpind, \Phi_\dpind)$ as in Eq.~\eqref{eq:shifted_polar_coords}. These provide coordinates in which the metric tensor, defined by Eq.~\eqref{eq:extended_metric}, is continuous across the DP horizon. The \emph{DP horizon} at $\ccoord_\dpind$ is formally defined as
\begin{equation}
    \mathcal{H}_q \equiv \left\{ (R_\dpind, \Phi_\dpind) \, \Big| \, R_\dpind = \tfrac{1}{2}, \, \Phi_\dpind \in [0, 2\pi) \right\},
\end{equation}
which is not contained in individual manifolds $\M^n$ or $\M^{n+1}$, but is well defined on $\M^{(n,n+1)}$. Its radius $\tfrac{1}{2}$ reflects the circumference $\CC=\pi$, but the bridge does not rely on this value: it follows from the two levels crossing along every line through the DP, and a~horizon of any other finite circumference would merely be drawn as a~circle of a~different radius.

Extending the construction of Eq.~\eqref{eq:connected_manifold_1D}, for manifolds connected by DPs between all levels from $\M^n$ up to $\M^{n+m}$, we introduce \emph{connected state manifold} (CSM) as
\begin{equation}
    \mathrm{CSM}^{(n, n+m)} = \left( \bigcup_{k=n}^{n+m} \overline{\M^{k}} \right) / \!\sim.
    \label{eq:CSM_definition}
\end{equation}

Usually, the Hamiltonian will not have DPs between all neighbouring eigenlevels. The energy gaps will split the manifold structure into disconnected CSM blocks, and we can collect all of them using the disjoint union
\begin{equation}
    \mathrm{CSM}^{(n_1, n_a)} \cup \mathrm{CSM}^{(n_b, n_c)} \cup \cdots.
    \label{eq:csm_split}
\end{equation}
Adiabatic driving on this large manifold, evolving the system between eigenlevels via DPs, can then happen only within the CSM of its origin.

\subsubsection{CSM properties}
The continuity of the metric on $\mathrm{CSM}^{(n,n+m)}$, including its values over the DP horizon \eqref{eq:extended_metric}, follows from the differentiability class $C^p$ of the underlying Hamiltonian $\H(\ccoord)$, as discussed in Sec.~\ref{sec:stretched_coords}. A~global atlas on CSM, in which the metric is of class $C^{p-2}$, can then be defined as a~collection of stretched coordinates around all DPs: $\{(R_\dpind, \Phi_\dpind)\}_{\dpind \in I_n\cup \dots \cup I_{n+m-1}}$, covering the whole manifold.

\begin{figure}[ht]
\centering
\resizebox{\linewidth}{!}{%
\begin{tikzpicture}[>=stealth, font=\small, every node/.style={inner sep=1pt}]

\def\rowA{0.3}
\def\rowB{-2.9}
\def\rowC{-6.1}
\def\rowBottom{-7.7}
\def\backopacity{0.2}

\node[font=\small\bfseries, align=center] at (0.3, 2.3) {Diabolic points};
\node[font=\small\bfseries, align=center] at (5.2, 2.3) {Topology};
\node[font=\small\bfseries, align=center] at (9, 2.3) {Graph repre.};

\draw[gray, thin] (-2.5, 1.9) -- (10.5, 1.9);
\draw[gray, thin] (-2.5, -1.3) -- (10.5, -1.3);
\draw[gray, thin] (-2.5, -4.5) -- (10.5, -4.5);
\draw[gray, thin] (3.0, 2.7) -- (3.0, \rowBottom);
\draw[gray, thin] (7.3, 2.7) -- (7.3, \rowBottom);

\node[font=\large\bfseries, anchor=north west] at (-2.3, 1.8) {a)};
\node[font=\large\bfseries, anchor=north west] at (-2.3, -1.4) {b)};
\node[font=\large\bfseries, anchor=north west] at (-2.3, -4.6) {c)};


\begin{scope}[shift={(0.3, \rowA)}] 
  \draw[thick] (0, -1.1) ellipse (0.753 and 0.185);
  \fill[white, opacity={1-\backopacity}] (0,0) -- (-0.753, 1.1) -- (0.753, 1.1) -- cycle;
  \fill[white, opacity={1-\backopacity}] (0,0) -- (-0.753, -1.1) -- (0.753, -1.1) -- cycle;
  \draw[thick] (0,0) -- (-0.753, 1.1);
  \draw[thick] (0,0) -- (0.753, 1.1);
  \draw[thick] (0, 1.1) ellipse (0.753 and 0.185);
  \draw[thick] (0,0) -- (-0.753, -1.1);
  \draw[thick] (0,0) -- (0.753, -1.1);
  \draw[red, very thick] (-0.18, 0.18) -- (0.18, -0.18);
  \draw[red, very thick] (-0.18, -0.18) -- (0.18, 0.18);
\end{scope}

\begin{scope}[shift={(5.4, \rowA)}]
    \draw[red, thick, dashed, opacity=\backopacity] (0.7, 0) arc (0:180:0.7 and 0.2);
  \draw[thick] (-0.7, -1.1) -- (-0.7, 1.1);
  \draw[thick] (0.7, -1.1) -- (0.7, 1.1);
  \draw[thick] (0, 1.1) ellipse (0.7 and 0.2);
  \draw[thick] (0.7, -1.1) arc (0:-180:0.7 and 0.2);
  \draw[thick, opacity=\backopacity] (0.7, -1.1) arc (0:180:0.7 and 0.2);
    \draw[red, thick, dashed] (0.7, 0) arc (0:-180:0.7 and 0.2);
  \node[right] at (0.7, 0.9) {$\M^1$};
  \node[right] at (0.7, -0.9) {$\M^0$};
\end{scope}

\begin{scope}[shift={(8.6, \rowA)}]
    \draw[red, thick, dashed] (0, 0.8) -- (0, -0.8);
    \fill (0, 0.8) circle (3pt);
    \fill (0, -0.8) circle (3pt);
    \node[right] at (0.1, 0.9) {$\M^1$};
    \node[right] at (0.1, -0.9) {$\M^0$};
\end{scope}


\begin{scope}[shift={(-0.55, \rowB)}]
  \draw[thick] (0, -0.95) ellipse (0.65 and 0.16);
  \fill[white, opacity={1-\backopacity}] (0,0) -- (-0.65, 0.95) -- (0.65, 0.95) -- cycle;
  \fill[white, opacity={1-\backopacity}] (0,0) -- (-0.65, -0.95) -- (0.65, -0.95) -- cycle;
  \draw[thick] (0,0) -- (-0.65, 0.95);
  \draw[thick] (0,0) -- (0.65, 0.95);
  \draw[thick] (0, 0.95) ellipse (0.65 and 0.16);
  \draw[thick] (0,0) -- (-0.65, -0.95);
  \draw[thick] (0,0) -- (0.65, -0.95);
  \draw[red, very thick] (-0.18, 0.18) -- (0.18, -0.18);
  \draw[red, very thick] (-0.18, -0.18) -- (0.18, 0.18);
\end{scope}
\begin{scope}[shift={(1.2, \rowB)}]
  \draw[thick, orange!80!black] (0, -0.95) ellipse (0.65 and 0.16);
  \fill[white, opacity={1-\backopacity}] (0,0) -- (-0.65, 0.95) -- (0.65, 0.95) -- cycle;
  \fill[white, opacity={1-\backopacity}] (0,0) -- (-0.65, -0.95) -- (0.65, -0.95) -- cycle;
  \draw[thick, orange!80!black] (0,0) -- (-0.65, 0.95);
  \draw[thick, orange!80!black] (0,0) -- (0.65, 0.95);
  \draw[thick, orange!80!black] (0, 0.95) ellipse (0.65 and 0.16);
  \draw[thick, orange!80!black] (0,0) -- (-0.65, -0.95);
  \draw[thick, orange!80!black] (0,0) -- (0.65, -0.95);
  \draw[MyBlue, very thick] (-0.18, 0.18) -- (0.18, -0.18);
  \draw[MyBlue, very thick] (-0.18, -0.18) -- (0.18, 0.18);
\end{scope}

\begin{scope}[shift={(5.4, \rowB)}]
    \draw[red, thick, dashed, opacity=\backopacity] (0.7, 0) arc (0:180:0.7 and 0.2);
  \draw[thick] (-0.7, -1.1) -- (-0.7, 1.1);
  \draw[thick] (0.7, -1.1) -- (0.7, 1.1);
  \draw[thick] (0, 1.1) ellipse (0.7 and 0.2);
  \draw[thick, opacity=\backopacity] (0.7, -1.1) arc (0:180:0.7 and 0.2);
  \draw[thick] (0.7, -1.1) arc (0:-180:0.7 and 0.2);
    \draw[red, thick, dashed] (0.7, 0) arc (0:-180:0.7 and 0.2);
  \draw[orange!80!black, thick] (0.48, 0.702) arc (90:-90:0.132 and 0.242);
  \draw[orange!80!black, thick] (0.48, -0.218) arc (90:-90:0.132 and 0.242);
  \draw[MyBlue, thick, dashed] (1.28, 0) arc (0:180:0.22 and {0.22/3.5});
  \fill[white, opacity={1-\backopacity}]
  (0.48, 0.702) .. controls (1.55, 0.702) and (1.55, -0.702) .. (0.48, -0.702)
  -- (0.48, -0.218) .. controls (0.96, -0.218) and (0.96, 0.218) .. (0.48, 0.218)
  -- cycle;
  \draw[orange!80!black, thick] (0.48, 0.702) arc (90:270:0.132 and 0.242);
  \draw[orange!80!black, thick] (0.48, -0.218) arc (90:270:0.132 and 0.242);
  \draw[MyBlue, thick, dashed] (1.28, 0) arc (0:-180:0.22 and {0.22/3.5});
  \draw[thick, orange!80!black]
  (0.48, 0.702) .. controls (1.55, 0.702) and (1.55, -0.702) .. (0.48, -0.702);
  \draw[thick, orange!80!black]
    (0.48, 0.218) .. controls (0.96, 0.218) and (0.96, -0.218) .. (0.48, -0.218);
  \node[right] at (0.7, 0.9) {$\M^1$};
  \node[right] at (0.7, -0.9) {$\M^0$};
\end{scope}

\begin{scope}[shift={(8.6, \rowB)}]
  \draw[red, thick, dashed] (0, 0.8) .. controls (-0.6, 0.35) and (-0.6, -0.35) .. (0, -0.8);
  \draw[MyBlue, thick, dashed] (0, 0.8) .. controls (0.6, 0.35) and (0.6, -0.35) .. (0, -0.8);
  \fill (0, -0.8) circle (3pt);
  \fill (0, 0.8) circle (3pt);
  \node[right] at (0.1, 0.9) {$\M^1$};
  \node[right] at (0.1, -0.9) {$\M^0$};
\end{scope}


\begin{scope}[shift={(0.4, \rowC+0.7)}]
  \begin{scope}[shift={(-0.95, -1.35)}]
    \draw[thick] (0, -0.8) ellipse (0.5 and 0.13);
    \fill[white, opacity={1-\backopacity}] (0,0) -- (-0.5, 0.8) -- (0.5, 0.8) -- cycle;
    \fill[white, opacity={1-\backopacity}] (0,0) -- (-0.5, -0.8) -- (0.5, -0.8) -- cycle;
    \draw[thick] (0,0) -- (-0.5, -0.8);
    \draw[thick] (0,0) -- (0.5, -0.8);
    \draw[thick] (0,0) -- (-0.5, 0.8);
    \draw[thick] (0,0) -- (0.5, 0.8);
    \draw[thick] (0, 0.8) ellipse (0.5 and 0.13);
    \draw[red, very thick] (-0.18, 0.18) -- (0.18, -0.18);
    \draw[red, very thick] (-0.18, -0.18) -- (0.18, 0.18);
  \end{scope}
  \begin{scope}[shift={(0.95, -1.35)}]
    \draw[thick, orange!80!black] (0, -0.8) ellipse (0.5 and 0.13);
    \fill[white, opacity={1-\backopacity}] (0,0) -- (-0.5, 0.8) -- (0.5, 0.8) -- cycle;
    \fill[white, opacity={1-\backopacity}] (0,0) -- (-0.5, -0.8) -- (0.5, -0.8) -- cycle;
    \draw[thick, orange!80!black] (0,0) -- (-0.5, -0.8);
    \draw[thick, orange!80!black] (0,0) -- (0.5, -0.8);
    \draw[thick, orange!80!black] (0,0) -- (-0.5, 0.8);
    \draw[thick, orange!80!black] (0,0) -- (0.5, 0.8);
    \draw[thick, orange!80!black] (0, 0.8) ellipse (0.5 and 0.13);
    \draw[MyBlue, very thick] (-0.18, 0.18) -- (0.18, -0.18);
    \draw[MyBlue, very thick] (-0.18, -0.18) -- (0.18, 0.18);
  \end{scope}
  \fill[white, opacity={1-\backopacity}] (-2.4, -1) -- (-1.6, -0.2) -- (2.4, -0.2) -- (1.6, -1) -- cycle;
  \draw[thick] (-2.4, -1) -- (-1.6, -0.2) -- (2.4, -0.2) -- (1.6, -1) -- cycle;
  \begin{scope}[shift={(0, -0.05)}]
    \draw[thick] (0, -0.6) ellipse (0.42 and 0.11);
    \fill[white, opacity={1-\backopacity}] (0,0) -- (-0.42, 0.6) -- (0.42, 0.6) -- cycle;
    \fill[white, opacity={1-\backopacity}] (0,0) -- (-0.42, -0.6) -- (0.42, -0.6) -- cycle;
    \draw[thick] (0,0) -- (-0.42, 0.6);
    \draw[thick] (0,0) -- (0.42, 0.6);
    \draw[thick] (0, 0.6) ellipse (0.42 and 0.11);
    \draw[thick] (0,0) -- (-0.42, -0.6);
    \draw[thick] (0,0) -- (0.42, -0.6);
    \draw[green!80!black, very thick] (-0.18, 0.18) -- (0.18, -0.18);
    \draw[green!80!black, very thick] (-0.18, -0.18) -- (0.18, 0.18);
  \end{scope}
  \begin{scope}[shift={(-0.55, 1)}]
  \draw[->, thick, gray] (-2.2, -3.2) -- (-2.2, -2.4) node[above] {$E$};
  \draw[->, thick, gray] (-2.2, -3.2) -- (-1.4, -3.2) node[right] {$\coord^1$};
  \draw[->, thick, gray] (-2.2, -3.2) -- (-1.75, -2.75) node[above right] {$\coord^2$};
  \end{scope}
\end{scope}

\begin{scope}[shift={(5.4, \rowC-0.1)}]
    \draw[red, thick, dashed, opacity=\backopacity] (0.7, 0) arc (0:180:0.7 and 0.2);
  \draw[thick] (-0.7, -1.1) -- (-0.7, 1.1);
  \draw[thick] (0.7, -1.1) -- (0.7, 1.1);
  \draw[thick] (0, 1.1) ellipse (0.7 and 0.2);
  \draw[thick, opacity=\backopacity] (0.7, -1.1) arc (0:180:0.7 and 0.2);
  \draw[thick] (0.7, -1.1) arc (0:-180:0.7 and 0.2);
    \draw[red, thick, dashed] (0.7, 0) arc (0:-180:0.7 and 0.2);
  \draw[orange!80!black, thick] (0.48, 0.702) arc (90:-90:0.132 and 0.242);
  \draw[MyBlue, thick, dashed] (1.28, 0) arc (0:180:0.22 and {0.22/3.5}); 
  \draw[orange!80!black, thick] (0.48, -0.218) arc (90:-90:0.132 and 0.242);
  \fill[white, opacity={1-\backopacity}]
  (0.48, 0.702) .. controls (1.55, 0.702) and (1.55, -0.702) .. (0.48, -0.702)
  -- (0.48, -0.218) .. controls (0.96, -0.218) and (0.96, 0.218) .. (0.48, 0.218)
  -- cycle;
  \draw[orange!80!black, thick] (0.48, -0.218) arc (90:270:0.132 and 0.242);
  \draw[orange!80!black, thick] (0.48, 0.702) arc (90:270:0.132 and 0.242);
    \draw[MyBlue, thick, dashed] (1.28, 0) arc (0:-180:0.22 and {0.22/3.5});
  \draw[thick, orange!80!black]
  (0.48, 0.702) .. controls (1.55, 0.702) and (1.55, -0.702) .. (0.48, -0.702);
  \draw[thick, orange!80!black]
    (0.48, 0.218) .. controls (0.96, 0.218) and (0.96, -0.218) .. (0.48, -0.218);
  \node[right] at (0.7, 0.9) {$\M^1$};
  \node[right] at (0.7, -0.9) {$\M^0$};

  \def\scy{0.55}
  \def\scx{-0.55}

  \def\sclen{0.9}
  \def\scr{0.28}
  \draw[thick] ({\scx}, \scy+\scr) arc (90:270:0.10 and \scr);
  \draw[thick] ({\scx}, \scy-\scr) arc (-90:90:0.10 and \scr);
    \draw[green!80!black, thick, dashed] ({\scx-0.45}, {\scy+\scr}) arc (90:270:0.10 and \scr);
  \fill[white, opacity={1-\backopacity}] (\scx, {\scy+\scr}) -- ({\scx-\sclen}, {\scy+\scr}) -- ({\scx-\sclen}, {\scy-\scr}) -- (\scx, {\scy-\scr}) -- cycle;
  \draw[thick] ({\scx-\sclen}, \scy) ellipse (0.10 and \scr);
  \draw[thick] (\scx, {\scy+\scr}) -- ({\scx-\sclen}, {\scy+\scr});
  \draw[thick] (\scx, {\scy-\scr}) -- ({\scx-\sclen}, {\scy-\scr});
    \draw[green!80!black, thick, dashed] ({\scx-0.45}, {\scy-\scr}) arc (-90:90:0.10 and \scr);
  \node[left] at ({\scx-\sclen-0.12}, \scy) {$\M^2$};
\end{scope}

\begin{scope}[shift={(8.6, \rowC)}]
    \draw[red, thick, dashed] (0, 0) .. controls (-0.5, -0.45) and (-0.5, -0.9) .. (0, -1.3);
    \draw[MyBlue, thick, dashed] (0, 0) .. controls (0.5, -0.45) and (0.5, -0.9) .. (0, -1.3);
    \draw[green!80!black, thick, dashed] (0, 1.15) -- (0, 0);
    \fill (0, 1.15) circle (3pt);
    \fill (0, 0) circle (3pt);
    \fill (0, -1.3) circle (3pt);
  \node[right] at (0.1, 1.15) {$\M^2$};
  \node[right] at (0.1, 0.1) {$\M^1$};
  \node[right] at (0.1, -1.35) {$\M^0$};
\end{scope}

\end{tikzpicture}%
}
\caption{Topological structure of CSMs. Left: schematic arrangement of spectrum in parametric space $\ccoord$; middle: the corresponding topology; right: the associated graph representation. (a) A~CSM connects two eigenlevels through a~single DP (red cross, dashed line). (b) Adding a~second DP between the same manifolds introduces a~handle on the infinite cylinder. (c) A~three-level CSM with two DPs between $\M^0$ and $\M^1$ and one DP between $\M^1$ and $\M^2$: the additional DP appears topologically as an extra cylinder attached to the surface in (b), modifying the topology only by adding an extra end.}
\label{fig:topology}
\end{figure}
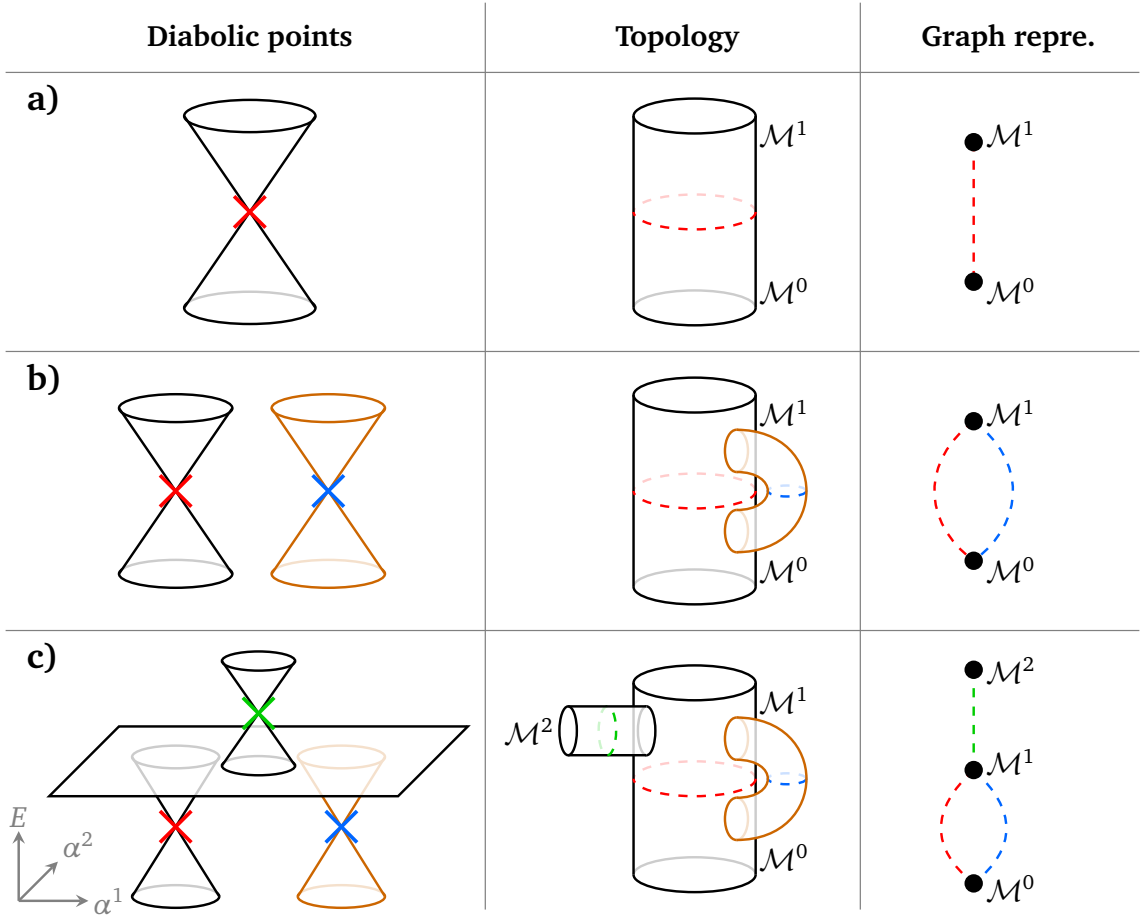

The topology of CSM grows complex with multiple DPs, as new loops and holes appear. We illustrate this on a~few simple cases, cf. Fig.~\ref{fig:topology}.
\begin{itemize}
    \item \textbf{Single DP on $\mathrm{CSM}^{(0, 1)}$:} has topology $S^1 \times \mathbb{R}$, a~cylinder with natural global coordinates $(\phi,E)$. The DP horizon serves as the cylinder's waist, separating $\M^0$ from $\M^1$. The cylinder's ends lie at infinity: the bottom circle corresponds to $R \to \infty$ ($r\rightarrow \infty$ on $\M^0$), the top circle corresponds to $R \to 0$ ($r\rightarrow -\infty$ on $\M^1$).
    
    \item \textbf{Two DPs on $\mathrm{CSM}^{(0, 1)}$:} A~second DP introduces another DP horizon. Topologically, this relates to an infinite cylinder with a~handle.

    \item \textbf{Three DPs on $\mathrm{CSM}^{(0, 2)}$:} Adding $\M^2$ glues an additional cylinder to the structure from the previous point.
\end{itemize}
The connectivity structure of a~CSM$^{(n,n+m)}$ also admits a~natural graph representation, shown in the third column of Fig.~\ref{fig:topology}. In this representation, vertices represent the eigenstate manifolds $\{\M^k\}$ and edges represent the DPs between them. This graph encodes which noncontractible adiabatic loops passing through their DPs exist in the CSM, independent of their precise realisation in parametric space.

The graph representation can be used to find the genus of the CSM structure, which corresponds to the number of independent cycles in the graph. Generally, the CSM has
\begin{equation}
    \text{genus } = -m + \sum_{k=n}^{n+m-1} |I_k|, \quad \text{number of ends} = m+1,
    \label{eq:genus}
\end{equation}
where $|I_k|$ marks the number of DPs between $\M^k$ and $\M^{k+1}$, and the ends (ideal boundary components) correspond to the spatial infinities of each manifold $\M^k$ in the sum. Because our space is non-compact, specifying the number of ends is important for the topological classification \cite{Richards63}.

Equation~\eqref{eq:genus} can be obtained recursively, adding one manifold and one DP at a~time. We start with a~single eigenstate manifold $\M^n$: a~plane without handles and with one end representing the spatial infinity. Connecting $\M^{n+1}$ via a~DP adds one end, the infinity of $\M^{n+1}$, but no handle. Every additional DP between the two manifolds adds a~handle, increasing the genus by $1$. Each additional manifold $\M^{k+1}$ is attached in the same way: its first DP to $\M^k$ adds only its end, and each extra DP adds a~handle. With $m+1$ manifolds connected by $\sum_k|I_k|$ DPs, of which $m$ are first DPs, this gives the genus $\sum_k|I_k|-m$ along with $m+1$ ends.

This classification is coordinate-independent and stable against continuous real perturbations $\H\mapsto \H+\delta \H$ that preserve the DP count between levels $n$ and $n+m$. Complex perturbations can move a~DP out of the 2D plane, changing it into an avoided crossing that no longer acts as a~bridge between eigenstate manifolds. Removing a~DP between levels $k$ and $k+1$ decreases the genus by one if $|I_k|\geq 2$; if it was the only one, the CSM splits into $\mathrm{CSM}^{(n,k)}$ and $\mathrm{CSM}^{(k+1,n+m)}$.

We have shown that the CSM provides more than a~regularisation of the metric near DPs. It expands the space accessible via adiabatic driving, as the trajectories are allowed to pass directly through DPs. These passages facilitate traversals between eigenstate manifolds and even permit the formation of loops that leverage the full potential of the nontrivial CSM topology, using multiple DPs as bridges.

\subsubsection{Compact parametric spaces}
\label{sec:compact_parametric}
The classification in Eq.~\eqref{eq:genus} assumes an unbounded parametric plane. For the parametric space being a~closed orientable surface~$P$ with $\text{genus}_P$ handles, such as the Brillouin-zone torus ($\text{genus}_P=1$), each $\M^k$ is a~copy of $P$, which has no ends. By the same argument as in the unbounded case, the first DP between $\M^k$ and $\M^{k+1}$ glues the two surfaces along a~single DP horizon, adding the $\text{genus}_P$ handles of $\M^{k+1}$, and every further DP adds one handle. The genus is therefore
\begin{equation}
    \text{genus}_\text{compact} = (m+1)\,\text{genus}_P-m+\sum_{k=n}^{n+m-1}|I_k|.
    \label{eq:genus_compact}
\end{equation}
A connected CSM over a~toroidal parametric space is thus a~closed surface of genus $1+\sum_k|I_k|$.

\subsection{Berry phase and nodal lines on the CSM}
\label{sec:nodal_lines}
Until now, we have focused mostly on the metric tensor, the real part of the quantum geometric tensor. However, analogous results hold even for its imaginary part, the Berry curvature, which defines the Berry phase acquired along an adiabatic driving protocol. Here we show how to compute this phase for arbitrary protocols, including those passing directly through DPs. 

We use the nodal-line approach, requiring the system to possess an effective $\PT$ symmetry. This is its main limitation relative to the gauge transformations approach of Ref.~\cite{Ju25}, yet it yields a~more comprehensive geometrical and topological understanding of the Berry phase.

\subsubsection{Nodal line theory on the CSM}
\label{sec:nodal_lines_theory}
We established in Sec.~\ref{sec:geometric_tensor} that DPs act as monopole-like sources of Berry curvature, producing quantised geometric flux as in Eq.~\eqref{eq:Berry_phase_quantization}. The Berry phase formula~\eqref{eq:Berry_phase} relies on the Stokes theorem, which requires a~smooth gauge---a continuous, single-valued choice of eigenstate phase---over the surface bounded by the path~$\Gamma$. However, the nonzero monopole flux emanating from a~DP means that no single smooth gauge can cover the entire parametric space surrounding it. The gauge must fail somewhere, and the locus of this failure is called a~\emph{nodal line}.

The local theory of these nodal lines, sometimes called Dirac strings~\cite{Ahn19}, was already developed in molecular physics, where DPs are known as conical intersections~\cite{Berry84Diabolical,Yarkony96}. The CSM extends this theory to a~global view, in which a~nodal line does not end at a~DP, as in Refs.~\cite{Mead92,Ahn19}, but passes through its horizon onto the adjacent manifold. The known rule that the Berry phase is given by the parity of nodal-line crossings~\cite{Mead92}, Eq.~\eqref{eq:gamma_b_crossings}, then holds for any closed path on the CSM, including paths through DPs.

Generally, a~nodal line is a~curve in parameter space along which a~chosen eigenstate gauge becomes singular: the eigenstate cannot be simultaneously single-valued and smooth everywhere~\cite{Mead79,Mead92,Zhao24}. It can be made single-valued either by a~cut (the nodal line), across which it changes sign, or by a~complex phase factor, which introduces a~vector potential~\cite{Mead79}. This is the direct analogue of the Dirac string in electromagnetism~\cite{Berry84}. For quantum eigenstates in a~3-parametric system, these singularities form one-dimensional line defects in parameter space.

For systems possessing an effective $\PT$ symmetry, eigenstates can be chosen real away from DPs, which reduces the full $U(1)$ gauge freedom to a~$\mathbb{Z}_2$ sign ambiguity. A~real eigenstate changes sign along a~loop around a~single DP~\cite{Herzberg63} and, conversely, a~sign change along a~loop implies a~degeneracy inside it~\cite{Herzberg63,LonguetHiggins75}, here a~DP. As a~consequence, the gauge singularities, which in a~general 3-parametric system form line defects in 3-dimensional parameter space, are naturally confined to the 2-parametric $\PT$-symmetric slice---precisely the space on which DPs and the CSM are defined. The nodal lines, therefore, appear as curves on the CSM itself, leading between DPs or extending to infinity.

More formally, any single-valued choice of phase on the parametric space can be made only patchwise; across the overlap of two patches, the states differ by a~sign transition function~\cite{Zhao24}.

While the nodal line position is gauge-dependent, the topological data it encodes (the Berry phase) is physical. A~continuous deformation of the chosen gauge, such as a~smooth variation of the projection vector $\bm w$ (Appendix~\ref{sec:appendix_AMM}), only displaces the nodal line without altering its endpoints~\cite{Mead79,Mead92}. On each manifold $\M^k$, this corresponds to "sliding" the branch cut across the surface while keeping its attachment points fixed. The quantisation $\gamma_B = k\pi \mod 2\pi$ established in Eq.~\eqref{eq:Berry_phase_quantization} can then be reinterpreted such that the Berry phase along any closed path $\Gamma$ depends only on the parity of nodal line crossings~\cite{Mead92,Louvet23}:
\begin{equation}
    \gamma_B(\Gamma) = \pi \cdot (\text{number of crossings}) \mod 2\pi.
    \label{eq:gamma_b_crossings}
\end{equation}

Consistency with the gauge structure imposes strict topological constraints on the configurations of these nodal lines. Topologically significant nodal lines must start and end at points where the gauge inherently fails: either at DPs or at parametric infinity~\cite{Mead92}. Furthermore, each DP must serve as an endpoint for an odd number of such nodal lines, since a~small loop around it reverses the sign of the eigenstate. These conditions apply only to topologically significant lines; it is also possible to observe \emph{closed nodal loops} that neither traverse a~DP nor extend to infinity. Such loops will always be traversed an even number of times by any continuous closed path on $\M^k$, meaning their Berry phase contribution is zero.

\begin{figure}
    \centering
    \includegraphics{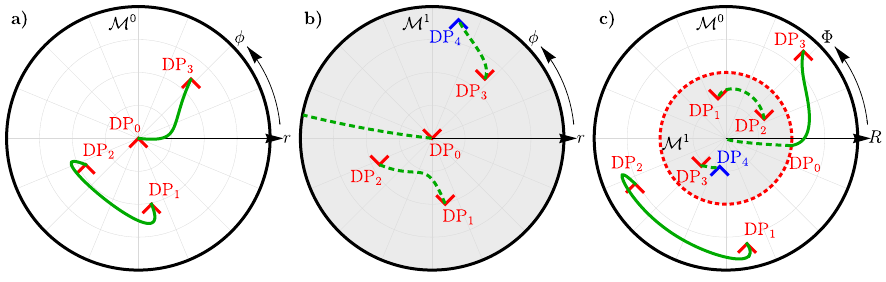}
    \caption{Demonstration of nodal lines for fictitious $\mathrm{CSM}^{(0,2)}$. (a) $\M^0$ in polar coordinates around $\mathrm{DP}_0$ with 4 DPs between $\M^0$ and $\M^1$(bottom halves of red crosses) and exemplary nodal lines between them (dashed, green). (b) $\M^1$ in the same polar coordinates, DPs are marked by upper halves of crosses, and possible nodal line continuation from $\M^0$ is shown. In addition, DP$_4$ between $\M^1$ and $\M^2$ is marked by a~blue half cross. (c) CSM in stretched coordinates around $\mathrm{DP}_0$ (red, dashed) with nodal line structure corresponding to the other two panels.}
    \label{fig:odd_vs_even_DPs}
\end{figure}

The parity of the total number of DPs determines the global nodal line structure and the resulting Berry phase. If a~single-state manifold contains an even number of DPs, all DPs can be paired with each other using nodal lines, as illustrated in Fig.~\ref{fig:odd_vs_even_DPs}(a). In this scenario, the Berry phase along a~large circular path enclosing all of them yields an even number of crossings, resulting in $\gamma_B = 0$. Conversely, if the number of DPs is odd, see Fig.~\ref{fig:odd_vs_even_DPs}(b), the enclosing path accumulates $\gamma_B = \pi$, and infinity itself must serve as an endpoint for an odd number of nodal lines to accommodate the unpaired DPs. In stretched coordinates, the infinity of $\M^{k+1}$ is mapped to the centre $R=0$. This point bears some resemblance to a~DP, since nodal lines can end or appear to pass through it. These nodal lines, however, do not have continuous derivatives at $R=0$; as one can imagine multiple nodal lines leading to the centre, similarly to the one drawn in Fig.~\ref{fig:odd_vs_even_DPs}(c).

On the CSM, DPs are no longer endpoints of nodal lines, since the continued eigenstate is smooth across the DP horizon (Sec.~\ref{sec:stretched_coords}). Each DP horizon is itself a~closed path, along which the continued eigenstate changes sign, so it has to be crossed by nodal lines an odd number of times. For any fixed gauge, the nodal lines thus traverse DPs onto the adjacent manifolds, and each nodal line on the CSM either closes on itself or runs off to the infinities at both ends. Equation~\eqref{eq:gamma_b_crossings} then gives the Berry phase of any closed path on the CSM, including paths passing through the DPs. Unlike on a~single-state manifold, closed nodal lines can thus be topologically significant, as the closed line through DP$_1$ and DP$_2$ in Fig.~\ref{fig:odd_vs_even_DPs}: a~closed path on $\M^0$ around DP$_1$ crosses it once. Likewise, a~closed nodal loop on $\M^k$ that encircles a~DP can be crossed once by a~closed path that enters it, passes through this DP and returns through another one. Only closed loops that encircle no DP can always be disregarded.

For CSM over the compact parametric space discussed in Sec.~\ref{sec:compact_parametric}, the nodal lines can no longer stretch to infinities, instead they have to close on themselves. A~closed nodal line can then also wind around a~handle of the parametric space, implying a~closed path along the perpendicular direction of this handle crosses it once, and such a~line cannot be disregarded even if it encircles no DP.

\subsubsection{Computation of nodal lines}
\label{sec:nodal_lines_practical}
The Berry phase along a~path $\Gamma$ is generally calculated by integrating the Berry connection, Eq.~\eqref{eq:Berry_phase}. This requires knowledge of the eigenvectors and their derivatives along the path. Practically, the wavefunction is usually obtained by numerical diagonalisation, where, for most algorithms, the gauge oscillates rapidly. This poses challenges for numerical differentiation and requires careful handling of boundary conditions at the path's ends.

When eigenvectors are computed with gauge fixed globally, the nodal line reveals itself as the locus where that gauge necessarily fails. This failure is hidden by standard numerical routines, as they pick an arbitrary local gauge and normalise the eigenvectors. To locate nodal lines systematically, we employ the \emph{adjugate matrix method} described in Appendix~\ref{sec:appendix_AMM}. This method fixes the gauge before computation and returns \emph{unnormalised eigenvectors} $\Psi$, with the zero vector magnitude indicating gauge-choice failure---a~nodal line. A~similar approach is used in Refs.~\cite{Mead92,Zhao24}.

We demonstrate it here on a~two-level system from Eq.~\eqref{eq:2level_hamiltonian_vector} with 3-dimensional parametric space $(h^1, h^2, h^3)$ and a~DP at the origin. The standard eigenvector patches fail along complementary half-lines in the $h^3$ direction through the origin, forming the nodal lines, as detailed in Ref.~\cite{Louvet23}. For example, the excited state can be written using two gauge choices as
\begin{equation}
    \ket{\psi_1} \propto \bm\Psi=\begin{pmatrix}
        |\bm h| + h^3 \\
        h^1 + ih^2
    \end{pmatrix}, \qquad
    \bm{\tilde\Psi}=\begin{pmatrix}
        h^1 - ih^2 \\
        |\bm h| - h^3
    \end{pmatrix}.
    \label{eq:2-level_eigenvectors}
\end{equation}
The unnormalized eigenvector vanishes along the line passing from the DP at the origin to infinity along the axis $\bm h = (0,0,h^3)^T$. For $\bm\Psi$ the nodal line goes along $h^3\leq 0$, for $\bm{\tilde\Psi}$ along $h^3 \geq 0$.

Restricting to the plane $h^2 = 0$ with a~real gauge, the 3-dimensional nodal line is projected to the 2-dimensional parametric space $(x,z)=(h^1,h^3)$, whose angular position rotates with the gauge choice. We parametrise the gauge by
\begin{equation}
    \bm w = (\cos\varphi, \sin\varphi)^T,
    \label{eq:2-level_gauge_parametrization}
\end{equation}
of which the magnitude is irrelevant (see Appendix~\ref{sec:appendix_AMM}). In this picture, the nodal line analytically continues through the DP, consistently with Sec.~\ref{sec:nodal_lines_theory}: its half-line on $\M^0$ and the opposite half-line on $\M^1$ form a~single line, which crosses the DP horizon once. Changing the gauge $\varphi$ rotates this line around the centre.

\begin{figure}
    \centering
    \includegraphics{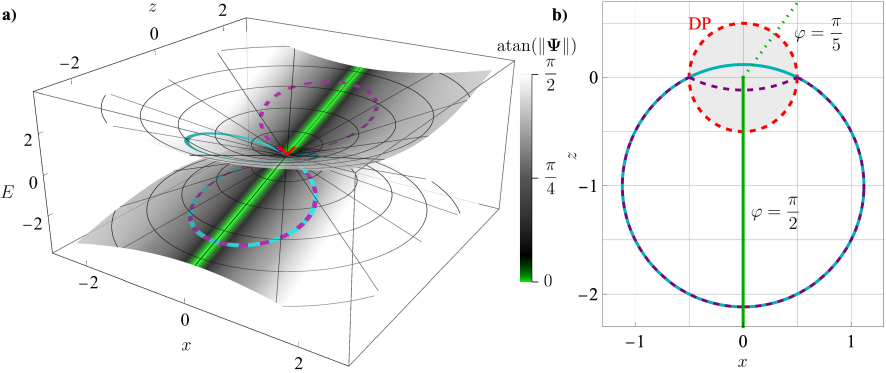}
    \caption{(a) A~conical structure of a~2-level system with $h^2=0$ and DP at $(0,0)$. The heatmap shows the magnitude of the unnormalised eigenvector, revealing the nodal line ($|\bm\Psi|=0$) passing through the DP. Two paths are drawn: a~double-circle (cyan) and a~figure-eight (purple, dashed). (b) The same scenario in stretched coordinates around the DP (red, dashed). Two gauges rotating the nodal line around the centre are shown: $\varphi\!=\!\pi/2$ (green, corresponding to panel~(a)) and $\varphi\!=\!\pi/5$ (green, dotted).}
    \label{fig:2-level_3D_energy}
\end{figure}

Figure~\ref{fig:2-level_3D_energy} shows the conical spectrum and a~heatmap of unnormalized eigenvector magnitude from Eq.~\eqref{eq:2-level_eigenvectors}, and two representative paths on the CSM: a~double-circle and a~figure-eight shaped. The double-circle path, which passes from $\M^0$ through the DP to $\M^1$ and back, crosses any topologically significant nodal line exactly once (independently of the gauge), yielding $\gamma_B = \pi$. The figure-eight crosses the nodal line zero or two times, as demonstrated in Fig.~\ref{fig:2-level_3D_energy}(b). Two gauges demonstrate this: $\varphi=\pi/5$ with no crossings, and $\varphi=\pi/2$ with two crossings, both leading to $\gamma_B = 0$ based on Eq.~\eqref{eq:gamma_b_crossings}. This is consistent with the analytic continuation picture: the double circle effectively encircles the DP once (the $\M^1$ arc being the analytically continued $\M^0$ from the opposite side), whereas the figure-eight traverses the same region twice without net encirclement, cf. Eq.~\eqref{eq:Berry_phase}.

\section{Applications in a~spin-1 system}
\label{sec:applications}

\subsection{System definition}
\label{sec:spin-1-system}
We demonstrate the CSM approach on a~minimal spin-1 system with zero-field splitting in an external magnetic field that hosts multiple DPs on a~single CSM.  Here, we use it only for illustration, without targeting a~specific application and experimental setting. This model describes the ${}^3 A_2$ ground electronic state of the nitrogen–vacancy centre in diamond~\cite{Doherty13,Rondin14,Gomez21}, whose effective Hamiltonian is ($\hbar=1$)
\begin{equation}
\hat{H} = D\hat{S}_z^2 + e(\hat{S}_x^2 - \hat{S}_y^2) + g\mu_B(B_x\hat{S}_x + B_y\hat{S}_y + B_z\hat{S}_z),
\end{equation}
for spin-1 operators $\hat{S}_{x,y,z}$, the axial zero-field splitting $D$, the transverse (rhombic) zero-field splitting parameter $e$ arising from non-axial crystal field components, the Landé g-factor $g\approx 2$, the Bohr magneton $\mu_B = 14 \;\mathrm{GHz/T}$, and the external magnetic field $\mathbf{B} = (B_x, B_y, B_z)$ measured in tesla.

To obtain a~two-parametric system $\ccoord = (B_x, B_z)$, we align the magnetic field along a~section $B_y=0$, and fix the $D,e$ parameters. We set the axial parameter from spin-spin dipolar interactions $D \approx 2.87$~GHz, and $e=0.5$~GHz. In the basis $\{\ket{m_s = +1}, \ket{m_s = 0}, \ket{m_s = -1}\}$, the Hamiltonian takes the form
\begin{equation}
\H(B_x, B_z) = \begin{pmatrix}
 B_z g \mu_B+D & \frac{B_x g \mu_B}{\sqrt{2}} & e \\
 \frac{B_x g \mu_B}{\sqrt{2}} & 0 & \frac{B_x g \mu_B}{\sqrt{2}} \\
 e & \frac{B_x g \mu_B}{\sqrt{2}} & D-B_z g \mu_B \\
\end{pmatrix}.
\label{eq:3-level_H}
\end{equation}
Its matrix representation is real, so it possesses an effective $\PT$ symmetry in the sense of Sec.~\ref{sec:geometric_tensor}. The system has two DPs between $\M^0$ and $\M^1$ at $\ccoord_0=(0,-z_0)$ and $\ccoord_1=(0,z_0)$, for $z_0=\sqrt{D^2 - e^2}/( g \mu_B)$, which for fixed parameters stated above yields $z_0=0.100933$~T, and it has an energy gap between $\M^1$ and $\M^2$, see Fig.~\ref{fig:3-level_spectrum}(a). For a~physical value $e>0$, the CSM structure splits into $\mathrm{CSM}^{(0,1)}$, as in Fig. \ref{fig:topology}(b), and a~separate $\M^2$ structure, i.e. $|I_0|=2$, $|I_1|=0$ according to Eq.~\eqref{eq:DP_set}.

For the extremal case $e=0$, the energy gap closes, making the CSM structure almost resemble the case in Fig. \ref{fig:topology}(c). This degenerate point, however, is not a~proper DP, breaking the second condition in Sec. \ref{sec:CSM_definition}. This \emph{non-diabolic degenerate point} is created by the interaction of the first and second eigenlevels in the $B_z$ direction, in which it behaves as a~DP with $E_2-E_1\propto B_z$; however, in the $B_x$ direction this interaction vanishes and only an avoided crossing between $\M^0$ and $\M^2$ is present, which accidentally touches the flat band $\M^1$ such that $E_2-E_1\propto B_x^2$. 

Our choice of $e=0.5$ GHz is convenient, as it makes the splitting $E_2-E_1|_{\bm B=\bm 0}=2e$ clearly resolved. While realistic diamond nanocrystal values span $e\lesssim 100$ kHz to a~few MHz \cite{Doherty13}, they produce the same qualitative features. For these values, the resulting energy splitting would be too small to be clearly visible in the figures that follow, so we adopt the larger value here for clarity.

\subsection{Numerical regularisation near diabolic points}
\label{sec:numerics_near_DPs}
In Cartesian coordinates $\ccoord=(B_x,B_z)$, the quantum metric tensor diverges near DPs: the energy gap closes and inflates the metric elements as $r^{-2}$ and the Christoffel symbols of the geodesic equation~\eqref{eq:geodesicEq} as $r^{-1}$ at a~distance $r$ from the DP, which makes numerical calculations of geometric quantities, such as geodesics, unreliable. This divergence comes from the coordinates, which collapse the DP horizon to a~point. This is exactly what the signed polar (and hence also the stretched) coordinates resolve. In them, the metric is smooth across the DP horizon, and the Christoffel symbols remain bounded away from the isolated directions in which the metric degenerates (Appendix~\ref{sec:aligned_coordinates}). A~different issue, which no choice of coordinates can change, is the conditioning of the eigenproblem itself: a~small perturbation $\delta\H$ coupling the two crossing levels, such as a~rounding error or an imprecision of the experimental parameters, changes the eigenstates by roughly $\|\delta\H\|/r$.

\begin{figure}[t]
\centering
    \centering 
    \includegraphics{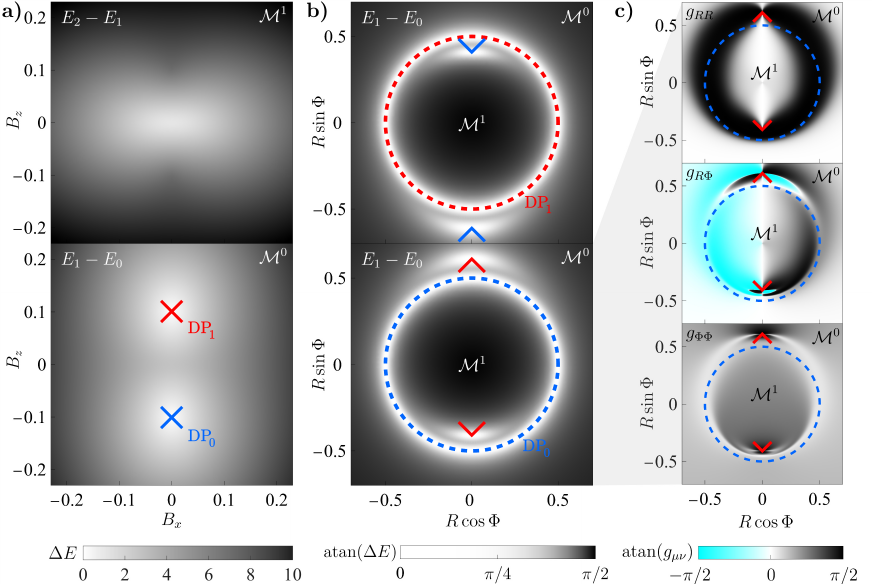}
    \caption{(a) Energy difference $\Delta E$ for system \eqref{eq:3-level_H}, along with DPs (red and blue crosses). (b) Energy difference $\arctan (\Delta E)$ in the stretched coordinates centred at marked DPs with corresponding colours. Inside and outside the DP horizon are marked by corresponding $n$th-level manifold along with DPs (half crosses) and DP horizons (dashed lines). (c) Metric tensor elements $\arctan g_{\mu\nu}$ in the stretched coordinates around $\mathrm{DP}_0$ are shown. $\wedge$ marks the lower half of the DP (when displayed on the lower manifold), $\vee$ marks the upper half (DP seen from the higher manifold).}
    \label{fig:3-level_spectrum}
\end{figure}

We illustrate the geometric regularisation on the spin-1 system~\eqref{eq:3-level_H} using stretched coordinates from Eq.~\eqref{eq:shifted_polar_coords} around DP coordinate $\ccoord_i$, such that $\ccoord -\ccoord_i = |r|(\cos\phi, \sin\phi)$. The parameters $(B_x,B_z)$ are not aligned in the sense of Eq.~\eqref{eq:two-level-alignment}, but by Eq.~\eqref{eq:c_is_pi} the horizon still has circumference $\CC=\pi$. Figure~\ref{fig:3-level_spectrum}(a) displays the energy difference $\Delta E$ between eigenlevels together with the two DPs on the ground state manifold. Panel~(b) displays $\Delta E$ in stretched coordinates centred at each DP (now spread over dashed horizons with corresponding colours): the interior region ($R<\tfrac{1}{2}$) now represents $\M^1$, and the exterior ($R>\tfrac{1}{2}$) represents $\M^0$, with each half cross marking the DP. Panel~(c) shows the metric tensor components $g_{\mu\nu}$ on the same chart around $\mathrm{DP}_0$ (blue).

This confirms that the singularity at the DP is coordinate removable, and the two sheets $\M^0$ and $\M^1$ are now connected into a~single manifold via the DP horizons, as the CSM theory dictates. This has practical consequences for geodesics, which can now cross the DP horizon $R=\tfrac{1}{2}$ with high numerical stability. Because geodesics are coordinate-invariant, results obtained in stretched coordinates translate directly back to Cartesian coordinates.

\subsection{Zero-determinant lines and geodesic fans}
\label{sec:singular_lines}
\begin{figure}
    \centering
    \includegraphics[width=\linewidth]{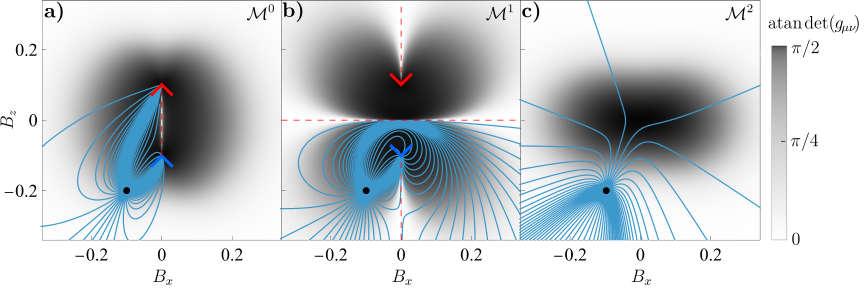}
    \caption{Metric tensor determinant calculated in Cartesian coordinates and geodesic fans for the model \eqref{eq:3-level_H}. Passages through lines $\det (g)=0$ (red, dashed) are structurally unstable, as demonstrated by deflecting geodesics. Geodesic paths start at an initial point $(B_x,B_z)|_{t=0}=(-0.1,-0.2)$ (black point) with velocity $(B_x',B_z')|_{t=0}=(\cos\theta,\sin\theta)$ uniformly distributed over the circle (60 geodesics in each panel). Each panel shows a~different eigenstate manifold marked in top right corner, and DPs are shown as in Fig.~\ref{fig:3-level_spectrum}.}
    \label{fig:singular_lines} 
\end{figure}

In the model \eqref{eq:3-level_H}, the vanishing linear terms $B_x$ and $B_z$ form \emph{zero-determinant lines}---one-dimensional curves in parametric space along which $\det(g)=0$. We show these in Fig.~\ref{fig:singular_lines} along with \emph{geodesic fans}, where multiple geodesics are evolved from a~single point with initial directions distributed uniformly. They are visibly attracted toward the DPs and deflected along the zero-determinant lines. On the $\M^2$, there are no zero-determinant lines, geodesics are only being repulsed by the avoided crossing between $E_1$ and $E_2$.

To see why some lines can act as barriers, consider the metric in Cartesian coordinates upon approaching such a~line. In our case, the metric component parallel to the line vanishes polynomially in the transverse distance. For instance, the transverse component $g_{B_zB_z}$ remains finite, while $g_{B_xB_x}\propto B_z^2$ vanishes near a~horizontal line, causing geodesics to be deflected along the zero-determinant line. For a~pedagogical example, see Appendix~\ref{sec:appendix_singular_lines}. These lines lead along the $\mathbb{Z}_2$ symmetries of the model: $B_z\to -B_z$, and $B_x\to -B_x$.

The only single-manifold geodesic that crosses a~zero-determinant line does so perpendicularly, a~condition that is infinitely sensitive to initial conditions: the set of initial conditions producing such a~crossing has measure zero in the space of initial directions. In parametric space, therefore, zero-determinant lines partition each manifold $\M^n$ into regions that are effectively disconnected from the viewpoint of geodesic adiabatic driving. Similar phenomenon of adiabatic obstructions was studied in~\cite{Souza16} and it underlies the geodesic impassability of quantum phase transitions in the thermodynamic limit~\cite{Kumar12, Strelecek25}.

\subsection{Geodesic shortcuts}
\label{sec:geodesic_shortcuts}
To the class of geodesics which exist on individual eigenstate manifolds, the CSM adds new ones, as the geodesics can now traverse DP horizons. Because the metric divergence at the DPs is shown to be removable by a~change of coordinates, the geodesic speed remains finite and well-defined even at DP crossings. We showcase one such case, in which the CSM geodesic is shorter than any single-manifold counterpart, though this is not guaranteed in general. The model presented is favourable to such a~comparison, because zero-determinant lines limit the reach of geodesics on each single-state manifold to a~subspace of the parametric space.

\begin{figure}[t]
    \includegraphics{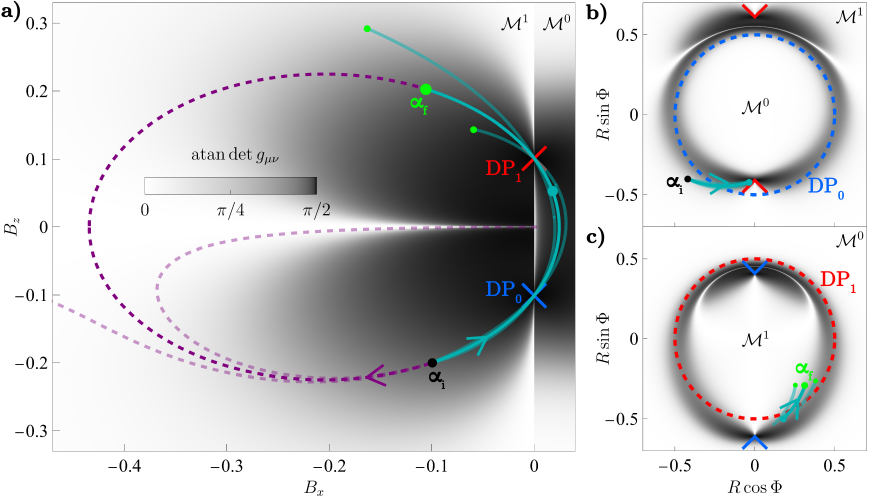}
    \caption{Comparison of an unstable single-state manifold geodesic (purple, dashed) with a~stable CSM geodesic (cyan). (a) The geodesic in Cartesian coordinates on the background of metric tensor, with $\det g^{(1)}_{\mu\nu}(B_x,B_z)$ for the $\M^1$ drawn on the $B_x<0$ half-plane, and of $\det g^{(0)}_{\mu\nu}(B_x,B_z)$ for the $B_x>0$ half-plane. All geodesic paths start on $\M^1$ at $\ccoord_i\equiv (B_x,B_z)|_{t=0}=(-0.1,-0.2)$ (black point) with unit velocity $\ccoord_i'=(\cos{(\theta_0)},\sin{(\theta_0)})$ with $\theta_0=0.4675$ for stable, and $\theta_0=\pi + 0.4675$ for unstable. The stable geodesic passes through $\mathrm{DP}_0$ (blue cross) to the $\M^0$ and back through $\mathrm{DP}_1$ (red cross) to the $\M^1$, and the evolution is aborted when it reaches $\ccoord_f=(-0.1,0.2)$ in time $t=0.5$ (green point). The unstable geodesic is evolved until it meets the point $\ccoord_f$. The first part of the stable geodesic evolution is calculated in stretched coordinates around $\mathrm{DP}_0$, as shown in panel (b) along with the perturbed geodesics. Similarly, the second part is calculated in stretched coordinates around $\mathrm{DP}_1$ as shown in panel (c). The transition is marked by a~cyan point and arrows mark the directions of time evolution. DPs are shown as in Fig.~\ref{fig:3-level_spectrum}.}
    \label{fig:geod_two_passages}
\end{figure}

The CSM framework opens a~concrete route to a~new kind of geodesics: by traversing a~DP horizon, the adiabatic path continues onto the adjacent manifold, and a~second DP may return it to the manifold of origin. Figure~\ref{fig:geod_two_passages} demonstrates this shortcut in the spin-1 system~\eqref{eq:3-level_H}. The geodesic (cyan) starts on $\M^1$ (black point) with initial velocity (cyan arrow) directed towards $\mathrm{DP}_0$, passes through it to $\M^0$, where it continues towards $\mathrm{DP}_1$ (red cross) through which it re-enters $\M^1$. Following Sec.~\ref{sec:stretched_coords}, we first compute time evolution in stretched coordinates around $\mathrm{DP}_0$, as shown in Fig.~\ref{fig:geod_two_passages}(b) and at a~sufficient distance from the $\mathrm{DP}_1$, the transition to the second local chart occurs (cyan point) to preserve numerical precision. The geodesic is then evolved further, as shown in Fig.~\ref{fig:geod_two_passages}(c), passing through $\mathrm{DP}_1$ back to $\M^1$ until it reaches time $t=0.5$ (green point), where the evolution is aborted.

We compare the CSM trajectory to another single-state manifold geodesic connecting the same endpoints. Evolving geodesics from $\ccoord_i$ on $\M^1$ in all directions, as in the geodesic fans of Fig.~\ref{fig:singular_lines}, we found that only the geodesic shown can cross to the half-plane $B_z>0$ on $\M^1$. The respective geometric lengths are summarised in Table~\ref{tab:geodesic_lengths}. This comparison of a~CSM geodesic passing through two DPs and a~single-state manifold geodesic shows that the CSM geodesic is indeed a~shortcut.

\begin{table}[t]
\centering
\begin{tabular}{r|c|c}
& CSM & $\M^1$ \\
\hline
trajectory & $\ccoord_i\rightarrow \mathrm{DP}_0 \rightarrow \mathrm{DP}_1 \rightarrow \ccoord_f$ & $\ccoord_i\rightarrow \text{zero-determinant line} \rightarrow\ccoord_f$\\
\hline
geometric length & $s_{\mathrm{CSM}}\approx 1.07$ & $s_{\M^1}\approx 2.08$ \\
\hline
initial condition stability & stable & unstable
\end{tabular}
\caption{Comparison of the CSM and single-state manifold geodesics.}
\label{tab:geodesic_lengths}
\end{table}

In addition, this shortcut is \emph{stable} with respect to initial conditions, while the single-state manifold geodesic is not. This is because the DP horizon is a~curve of geometric length $\pi$, on which the metric is non-degenerate except in isolated directions; a~geodesic that crosses it away from these directions keeps crossing it when its initial direction is perturbed, whereas crossing a~zero-determinant line, on which the metric degenerates, requires a~fine-tuned initial direction. Figure~\ref{fig:geod_two_passages}(a) shows, alongside each of the two geodesics, two companion trajectories with initial angle perturbed by $\Delta\theta_0$. For the CSM shortcut, perturbations of $\Delta\theta_0=\pm 0.1$ still arrive to the neighbourhood of the green endpoint. The evolution time for these geodesics is set such that they travel the same distance $s_{\mathrm{CSM}}$. For the single-manifold geodesic, even an infinitesimal perturbation already produces macroscopic deflection along the zero-determinant line $B_z=0$. For demonstration we show $\Delta\theta_0=\pm 0.01~\mathrm{rad}$.\footnote{Because of the instability with respect to the initial condition, the unstable geodesic was calculated by optimisation of geodesic evolution from $(B_x,B_z)|_{t=0}=(x, 10^{-3})$ in the $-B_z$ direction with unit velocity, for which the geodesic passes through the $(-0.1,-0.2)$ point. It was then mirrored to the $B_z$ initial direction, using the symmetry $B_z\to-B_z$ of the model. This is permissible because the shape of the geodesic and the distance travelled are independent of its orientation and speed. Such a~path approximates the true geodesic within a~small error acceptable for our analysis, for $\theta_0$ the error is smaller than $10^{-5}$.}

\subsection{Nodal lines}
\label{sec:nodal_lines_spin1}
Just as geodesics pass continuously through DPs, the nodal lines defined in Sec.~\ref{sec:nodal_lines}, pass as well. For the 3-level spin-1 Hamiltonian \eqref{eq:3-level_H}, we compute nodal lines using the gauge-fixed construction described in Sec.~\ref{sec:nodal_lines_practical} (see also Appendix~\ref{sec:appendix_AMM}). Only the direction of the real gauge vector $\bm w\in\R^N$ matters (Appendix~\ref{sec:appendix_AMM}), so it is parametrised by $N-1$ angles: a~single angle for the two-level system in Eq.~\eqref{eq:2-level_gauge_parametrization}, and two spherical angles $\chi,\varphi$ for the present $N=3$ system: \begin{equation}
    \bm w = (\sin\chi\cos\varphi, \sin\chi\sin\varphi, \cos\chi).    
\end{equation}
We show the nodal-line structure for two representative gauges. 

\begin{figure}
    \centering
    \includegraphics{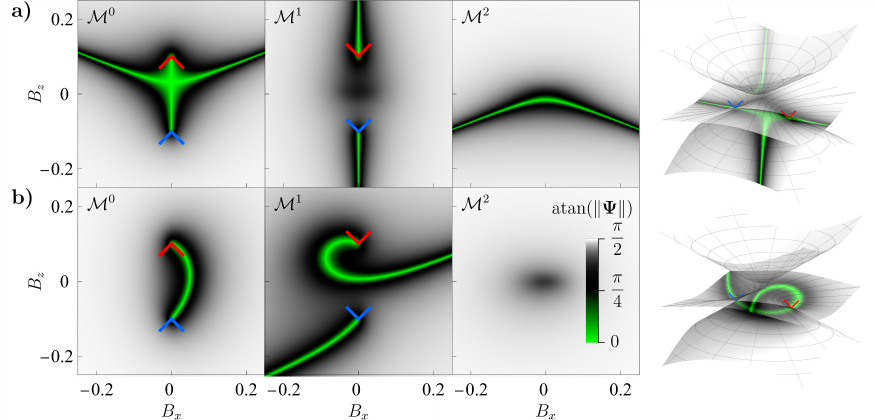}
    \caption{Nodal lines of the 3-level system~\eqref{eq:3-level_H} for gauge choices: (a) for $(\chi,\varphi)=(2,0)$, and (b) for $(\chi,\varphi)=(1,0.4)$. Columns correspond to manifolds $\M^k$. Both are visualised in the $(B_x,B_z,E)$ space on the right, as in Fig.~\ref{fig:illustration}. DPs are marked by crosses (and half crosses as seen from individual manifolds).}
    \label{fig:3-level_3D_energy}
\end{figure}

In the first, we set $(\chi,\varphi) = (2, 0)$, see Fig.~\ref{fig:3-level_3D_energy}(a); the nodal line extends from infinity $(B_x, B_z) = (0, \infty)$ on $\M^1$, passes through $\mathrm{DP}_1$ to $\M^0$, continues to $\mathrm{DP}_0$, through which it passes back to $\M^1$. The other two nodal lines are confined to $\M^0$ and $\M^2$, passing between the DPs from infinity to infinity. 

In the second gauge, Fig.~\ref{fig:3-level_3D_energy}(b), $(\chi,\varphi) = (1, 0.4)$, the nodal line extends from infinity on $\M^1$, passes through $\mathrm{DP}_0$ to $\M^0$, returns through $\mathrm{DP}_1$ to $\M^1$, and continues to infinity. There is no other nodal line in this gauge passing near the DPs.

If we were to close the energy gap by setting $e=0$, the nodal line would be allowed to continue through the non-diabolic degenerate point at $\bm B=\bm 0$ between $\M^1$ and $\M^2$ in the $B_z$ direction due to the linear crossing present: the degenerate point behaves the same as a~proper DP in this direction. This is, however, forbidden in the $B_x$ direction, where the energy difference is quadratic, and the eigenvector smoothly continues on the same manifold. This is exactly what would happen in the second gauge, where the nodal line passing through $\bm B=0$ would continue on $\M^1$ instead of passing to $\M^2$. This shows that non-diabolic degenerate points do not act as bridges between different manifolds, and the CSM would need advanced modifications to accommodate them.

\section{Conclusion}
\label{sec:conclusion}
We introduced the connected state manifold (CSM), a~unified geometric framework that resolves the singular behaviour of the metric tensor at DPs. Each DP, which traditionally leads to singularities in the geometry of its adjacent eigenstate manifolds $\M^n$ and $\M^{n+1}$, is promoted to a~circular boundary: the DP horizon. This boundary between eigenstate manifolds has a~nonzero geometric circumference, across which the two manifolds are glued into a~single surface. With the signed polar coordinates, the metric extends smoothly across the DP horizon, where, for generic couplings to the other levels, it is non-degenerate except in isolated directions. The CSM thus emerges as the natural geometric object on which geometric tensor, geodesics, and Berry connection can be analysed across the DPs.

We demonstrated three concrete advances on a~spin-1 system with multiple DPs. The first advance addresses an obstruction around the DPs: in Cartesian parametrisation, the closing energy gap inflates the components of the metric tensor and the coefficients of the geodesic equation in the neighbourhood of the DP. Signed polar coordinates around individual DPs remove these divergences, so that geodesics can be integrated through DPs, and the stretched coordinates then provide the charts in which the CSM is constructed and visualised.

The second advance opens an entirely new class of geodesics by permitting trajectories to traverse the full CSM rather than being confined to individual eigenstate manifolds, as they can now cross DP horizons. Whether these new CSM trajectories are geometrically shorter than the single-manifold counterparts depends on the specific Hamiltonian and endpoints, and is not guaranteed by the construction. We demonstrated this advantage in the spin-1 model, where a~CSM geodesic passing through two DPs in succession is about half as long as the single-manifold geodesic connecting the same endpoints and, unlike it, stable. We emphasise that the stability addressed here is the trajectory's sensitivity to its initial direction, which is conceptually distinct from the parameter precision required to drive the system through a~DP in a~standard adiabatic protocol. The geodesic formulation does not eliminate parameter sensitivity; it reformulates the trajectory as a~coordinate-invariant curve, transferring the question of stability from the schedule $\ccoord(t)$ to the geodesic initial condition $\ccoord'(t=0)$.

The third advance concerns the antisymmetric part of the quantum geometric tensor---the Berry curvature. While the CSM was constructed primarily with the symmetric part of the geometric tensor in mind (the metric tensor), the same continuity arguments extend to the Berry curvature, so that nodal lines of any chosen real gauge become curves on the CSM that pass through DPs or extend to infinity. The Berry phase accumulated by the adiabatic driving along a~closed path (including paths through DPs themselves) reduces to a~parity count of nodal-line crossings on this connected surface. The topological constraints that the nodal-line configuration must satisfy can be directly visualised on the CSM rather than being scattered across disconnected eigenstate manifolds.

Several extensions and limitations are worth highlighting. The CSM construction requires linear energy splitting at degeneracies, as in Eq.~\eqref{eq:DP_definition_endif}. Relaxing this condition leads to a~non-diabolic degenerate point, at which the metric need not extend continuously to the horizon, and the passage through it may be restricted along some directions or entirely, as discussed for the spin-1 model with a~specific parameter setup. Our topological classification of the CSM further assumes that the DPs occur between adjacent eigenlevels, and that the QGT is Abelian (the non-Abelian case lies outside the present framework and would require its own treatment). Such manifolds go beyond our CSM formulation. A~physical example beyond these assumptions is the rotating shallow-water model~\cite{Ganeshan26}, in which all three bands meet at a~single point of codimension three without an effective $\PT$ symmetry. We also discussed how the topology changes when the parametric space is compact, such as the Brillouin zone. There the CSM then becomes a~closed surface, without the ends of the planar case and with additional handles from the parametric space. In this case, the genus could also be obtained from the integrated curvature of the metric via the Gauss--Bonnet theorem, as done for ground-state manifolds without the DPs in Ref.~\cite{Kolodrubetz13}. The theorem assumes a~non-degenerate metric, so zero-determinant lines and other sets on which the metric degenerates require separate treatment. A~final comment concerns the role of $\PT$ symmetry. It is essential only for the Berry-phase analysis, where the reduction of the gauge group from $U(1)$ to $\mathbb{Z}_2$ confines nodal lines to the two-parameter slice and quantises the Berry phase to multiples of $\pi$. In the rest of the article, we use it purely for convenience, as it ensures that DPs appear generically in two-parameter systems of interest.

The CSM provides new tools for any parametric quantum system with DPs, and admits development along several directions. These include triple crossings and the codimension-three degeneracies of general Hermitian systems. Other directions are the mathematical study of systems with broken conditions required for CSM construction, or finding the correspondence between general relativity wormholes and CSM bridges: specific Hamiltonians $\hat H(r,\phi)$ can produce CSM with a~metric that matches some known two-dimensional wormhole solution. Viewed in this light, DPs cease to be obstructions to a~geometric description and emerge as the new elements enriching the structure of eigenstate manifolds.

\section*{Acknowledgements} 
JS thanks Marin Bukov, Zuo Wang, Diego Liska and Luk\'a\v{s} Honsa for valuable discussions.

\paragraph{Funding information}
We acknowledge the financial support of the Czech Science Foundation grant no. 25-16056S, the Charles University Research Center of Excellence UNCE 24/SCI/016, and the Grant Agency of Charles University GAUK~207723.

\paragraph{Data availability}
The code used for numerical analysis in this work is openly available on Zenodo at \doi{10.5281/zenodo.20340521}.

\begin{appendix}
\numberwithin{equation}{section}

\section{Metric tensor around the DP (general multi-level case)}
\label{sec:appendix_SPT}
In this section, we derive expressions for the geometric tensor in the limit $r\to 0$ for the Hamiltonian~\eqref{eq:H_expansion_DP} with spectrum as in Eq.~\eqref{eq:spectrum_linearization}. We use the stationary perturbation theory (SPT) adapted for a~$\phi$-dependent perturbation. Because the unperturbed Hamiltonian $\H_0$ is degenerate at the DP at $r=0$, determining the first-order wavefunction corrections within the degenerate subspace requires expanding the Schr\"odinger equation to the second order in $r$.

We proceed directly from the Taylor expansion of $\H$ with $r$-independent coefficients,
\begin{equation}
    \H(r,\phi) = \H_0 + r\,\H_I^{\{1\}}(\phi) + r^2\,\H_I^{\{2\}}(\phi) + \O(r^3),
    \label{eq:H_Taylor}
\end{equation}
where $\H_0 \equiv \H(r=0)$, $\H_I^{\{1\}}(\phi) = \bm{f^{\{1\}}}(\phi)\cdot\hat{\bm{\lambda}}$ is the first-order term, and $\H_I^{\{2\}}(\phi)$ is the second-order coefficient from Eq.~\eqref{eq:H_expansion_DP}. We similarly expand
\begin{align}
    E_n(r,\phi) &= E_n^{\{0\}} + rE_n^{\{1\}}(\phi) + r^2 E_n^{\{2\}}(\phi) + \O(r^3),
    \label{eq:E_expansion}\\
    \ket{\psi_n(r,\phi)} &= \ket{\psi_n^{\{0\}}(\phi)} + r\ket{\psi_n^{\{1\}}(\phi)} + r^2\ket{\psi_n^{\{2\}}(\phi)} + \O(r^3).
    \label{eq:psi_expansion}
\end{align}
Substituting Eqs.~\eqref{eq:H_Taylor}--\eqref{eq:psi_expansion} into $\H\ket{\psi_n} = E_n\ket{\psi_n}$ and matching powers of $r$ gives the hierarchy (omitting the functional dependences)
\begin{align}
    \O(r^0):&\quad \H_0\ket{\psi_n^{\{0\}}} = E_n^{\{0\}}\ket{\psi_n^{\{0\}}}, \label{eq:SPT_order0}\\[4pt]
    \O(r^1):&\quad (\H_0 - E_n^{\{0\}})\ket{\psi_n^{\{1\}}} = (E_n^{\{1\}} - \H_I^{\{1\}})\ket{\psi_n^{\{0\}}}, \label{eq:SPT_order1}\\[4pt]
    \O(r^2):&\quad (\H_0 - E_n^{\{0\}})\ket{\psi_n^{\{2\}}} = (E_n^{\{1\}} - \H_I^{\{1\}})\ket{\psi_n^{\{1\}}} + (E_n^{\{2\}} - \H_I^{\{2\}})\ket{\psi_n^{\{0\}}}. \label{eq:SPT_order2}
\end{align}
In Eq.~\eqref{eq:SPT_order0}, the zeroth-order states $\ket{\psi_0^{\{0\}}}$ and $\ket{\psi_1^{\{0\}}}$ span the degenerate subspace but are otherwise undetermined at this order. This equation only sets the zeroth order energies $E_n^{\{0\}} = \braket{\psi_n^{\{0\}}|\H_0|\psi_n^{\{0\}}}$. For Eq.~\eqref{eq:SPT_order1} to yield a~solution $\ket{\psi_0^{\{1\}}}$, the right-hand side must be orthogonal to the degenerate subspace, which leads to the condition
\begin{equation}
    \braket{\psi_1^{\{0\}}|\H_I^{\{1\}}|\psi_0^{\{0\}}} = 0.
    \label{eq:good_state}
\end{equation}
The first-order energies are then $E_n^{\{1\}} = \braket{\psi_n^{\{0\}}|\H_I^{\{1\}}|\psi_n^{\{0\}}}$ and splitting $\Delta_1 \equiv E_1^{\{1\}} - E_0^{\{1\}} > 0$.
Projecting Eq.~\eqref{eq:SPT_order1} for $n=0$ onto $\bra{\psi_k^{\{0\}}}$ gives us
\begin{equation}
    \braket{\psi_k^{\{0\}}|\psi_0^{\{1\}}} = -\frac{\braket{\psi_k^{\{0\}}|\H_I^{\{1\}}|\psi_0^{\{0\}}}}{\Delta_k}, \quad k \geq 2.
    \label{eq:psi1_outside}
\end{equation}
Projection onto $\bra{\psi_1^{\{0\}}}$ yields a~trivially satisfied equation, and we need to evaluate the second-order terms to determine $c \equiv \braket{\psi_1^{\{0\}}|\psi_0^{\{1\}}}$.

From the second order \eqref{eq:SPT_order2}, projecting onto $\bra{\psi_1^{\{0\}}}$ for $n=0$, we get
\begin{equation}
    0 = E_0^{\{1\}}\,c - \braket{\psi_1^{\{0\}}|\H_I^{\{1\}}|\psi_0^{\{1\}}} - \braket{\psi_1^{\{0\}}|\H_I^{\{2\}}|\psi_0^{\{0\}}}.
    \label{eq:SPT2_constraint}
\end{equation}
We expand $\ket{\psi_0^{\{1\}}}$ in the zeroth-order eigenbasis
\begin{equation}
    \begin{split}
        \ket{\psi_0^{\{1\}}} &= \underbrace{\braket{\psi_0^{\{0\}}|\psi_0^{\{1\}}}}_{=\,0}\ket{\psi_0^{\{0\}}} 
        + \underbrace{\braket{\psi_1^{\{0\}}|\psi_0^{\{1\}}}}_{\equiv\, c}\ket{\psi_1^{\{0\}}} 
        + \sum_{k=2}^{N-1}\underbrace{\braket{\psi_k^{\{0\}}|\psi_0^{\{1\}}}}_{\text{Eq.~}\eqref{eq:psi1_outside}}\ket{\psi_k^{\{0\}}} \\
        &= c\ket{\psi_1^{\{0\}}} - \sum_{k=2}^{N-1}\frac{\braket{\psi_k^{\{0\}}|\H_I^{\{1\}}|\psi_0^{\{0\}}}}{\Delta_k}\ket{\psi_k^{\{0\}}},
        \label{eq:psi1_expansion}
    \end{split}
\end{equation}
The first coefficient can be shown to vanish when we insert Eq.~\eqref{eq:psi_expansion} into the normalisation condition $\braket{\psi_0(r)|\psi_0(r)}=1$. At linear $r$ term, this leads to $2\,\Re\braket{\psi_0^{\{0\}}|\psi_0^{\{1\}}}=0$, where the remaining imaginary part corresponds to a~global phase freedom and can be fixed by requiring
\begin{equation}
    \braket{\psi_0^{\{0\}}|\psi_0^{\{1\}}}=0.
    \label{eq:psi1_gauge} 
\end{equation}

Applying $\bra{\psi_1^{\{0\}}}\H_I^{\{1\}}$ to Eq.~\eqref{eq:psi1_expansion} gives
\begin{equation}
    \braket{\psi_1^{\{0\}}|\H_I^{\{1\}}|\psi_0^{\{1\}}} = c E_1^{\{1\}} - \sum_{k=2}^{N-1}\frac{\braket{\psi_k^{\{0\}}|\H_I^{\{1\}}|\psi_0^{\{0\}}}\braket{\psi_1^{\{0\}}|\H_I^{\{1\}}|\psi_k^{\{0\}}}}{\Delta_k}.
    \label{eq:H_psi1_expansion}
\end{equation}
Substituting into Eq.~\eqref{eq:SPT2_constraint} and using $\Delta_1\equiv E_1^{\{1\}}-E_0^{\{1\}}$ we get
\begin{equation}
    c \equiv \braket{\psi_1^{\{0\}}|\psi_0^{\{1\}}} = -\frac{\braket{\psi_1^{\{0\}}|\H_I^{\{2\}}|\psi_0^{\{0\}}}}{\Delta_1} + \frac{1}{\Delta_1}\sum_{k=2}^{N-1}\frac{\braket{\psi_1^{\{0\}}|\H_I^{\{1\}}|\psi_k^{\{0\}}}\braket{\psi_k^{\{0\}}|\H_I^{\{1\}}|\psi_0^{\{0\}}}}{\Delta_k}.
    \label{eq:c1_general}
\end{equation}
The first term shows that the second-order Hamiltonian coefficient $\H_I^{\{2\}}$ generically contributes to $c$ and therefore to $\tilde{g}_{rr}^{(0)}$. It also explains why first-order perturbation theory alone fails to derive correct $g_{rr}$, as we illustrate later for a~3-level system in Fig.~\ref{fig:grrProblem}.

We now use these results to derive each component of the geometric tensor in the limit $r\to 0$.

\subsection{Components of the metric tensor}
\label{sec:appendix_metric_components}
\paragraph{Angular component $\tilde{T}_{\phi\phi}^{(0)}$:}
Here only the first-order expansion $\partial_\phi\H = r\,\partial_\phi\H_I^{\{1\}} + \O(r^2)$ and $E_1 - E_0 = r\Delta_1 + \O(r^2)$ suffices. Inserting into Eq.~\eqref{eq:QGT_definition} we get
\begin{equation}
    \begin{split}
        \tilde T_{\phi\phi}^{(0)} &= \lim_{r\to 0} \sum_{j=1}^{N-1} \frac{\left|\braket{\psi_j|\partial_\phi \H|\psi_0}\right|^2}{\left(E_j - E_0\right)^2} \\
        &= \lim_{r\to 0} \frac{\left|\braket{\psi_1|\partial_\phi \H_I^{\{1\}}|\psi_0} + \O(r)\right|^2}{\left(\Delta_1 + \O(r)\right)^2}
        + \sum_{j=2}^{N-1}\frac{r^2\left|\braket{\psi_j^{\{0\}}|\partial_\phi \H_I^{\{1\}}|\psi_0^{\{0\}}}\right|^2 + \O(r^3)}{\left(\Delta_j + \O(r)\right)^2}\\
        &= \frac{\left|\braket{\psi_1^{\{0\}}|\partial_\phi \H_I^{\{1\}}|\psi_0^{\{0\}}}\right|^2}{\Delta_1^2}.
    \end{split}
    \label{eq:Tphiphi_result}
\end{equation}
$\tilde{T}_{\phi\phi}^{(0)}$ is therefore determined entirely by the degenerate pair and is real, leading to $\tilde{g}_{\phi\phi}^{(0)} = \tilde{T}_{\phi\phi}^{(0)}$.

\paragraph{Radial component $\tilde T_{rr}^{(0)}$:}
Inserting series \eqref{eq:psi_expansion} into the first form of Eq.~\eqref{eq:QGT_definition}, we get
\begin{equation}
    \tilde T_{rr}^{(0)}  = \braket{\psi_0^{\{1\}}|\psi_0^{\{1\}}}-|\underbrace{\braket{\psi_0^{\{1\}}|\psi_0^{\{0\}}}}_{=0 \text{ from}\;\eqref{eq:psi1_gauge}}|^2= \braket{\psi_0^{\{1\}}|\psi_0^{\{1\}}} \underbrace{=}_{\eqref{eq:psi1_expansion}} |c|^2 + \sum_{k=2}^{N-1}\frac{\left|\braket{\psi_k^{\{0\}}|\H_I^{\{1\}}|\psi_0^{\{0\}}}\right|^2}{\Delta_k^2},
    \label{eq:Trr_result}
\end{equation}
where $c$ is determined by Eq.~\eqref{eq:c1_general}. The radial metric element is therefore non-negative, real, and depends explicitly on the quadratic Hamiltonian term $\H_I^{\{2\}}$ and all eigenlevels of the system.

\paragraph{Off-diagonal component $\tilde T_{r\phi}^{(0)}$:}
Differentiating $\H\ket{\psi_0} = E_0\ket{\psi_0}$ with respect to $r$ and projecting onto $\bra{\psi_j}$, $j\neq 0$, yields the Hellmann--Feynman identity
\begin{equation}
    \braket{\psi_j|\partial_r\psi_0}=\frac{\braket{\psi_j|\partial_r\H|\psi_0}}{E_0-E_j}.
\end{equation}
Applying this to the radial factor in the summation form of Eq.~\eqref{eq:QGT_definition} gives
\begin{equation}
    T_{r\phi}^{(0)} = \sum_{j\neq 0}
    \frac{\braket{\partial_r\psi_0|\psi_j}\braket{\psi_j|\partial_\phi\H|\psi_0}}{E_0-E_j}.
    \label{eq:Trphi_summation}
\end{equation}
Using $\ket{\partial_r\psi_0}\to\ket{\psi_0^{\{1\}}}$, $\partial_\phi\H = r\,\partial_\phi\H_I^{\{1\}}(\phi)+\mathcal{O}(r^2)$ from Eq.~\eqref{eq:H_Taylor}, each term in the sum behaves as
\begin{equation}
    \frac{\braket{\psi_0^{\{1\}}|\psi_j^{\{0\}}}\,
    \braket{\psi_j^{\{0\}}|r\,\partial_\phi\H_I^{\{1\}}|\psi_0^{\{0\}}}+\mathcal{O}(r^2)}{E_0-E_j}.
\end{equation}
We distinguish two cases in the limit: the $j\geq 2$ terms have $E_j - E_0 \to \Delta_j = \mathrm{const}$, so the factor $r$ from $\partial_\phi\H$ makes them vanish. For $j=1$ the denominator $E_1-E_0 = r\Delta_1+\mathcal{O}(r^2)$ also carries a~factor of $r$, which exactly cancels, leaving a~finite result. Therefore, only the degenerate pair survives:
\begin{equation}
    \tilde T_{r\phi}^{(0)}
    = \lim_{r\to 0}\left(-\frac{\braket{\psi_0^{\{1\}}|\psi_1^{\{0\}}}
    \cdot r\braket{\psi_1^{\{0\}}|\partial_\phi\H_I^{\{1\}}|\psi_0^{\{0\}}}
    + \mathcal{O}(r^2)}{r\Delta_1+\mathcal{O}(r^2)}\right)
    = -\frac{c^*\,
    \braket{\psi_1^{\{0\}}|\partial_\phi\H_I^{\{1\}}|\psi_0^{\{0\}}}}{\Delta_1},
    \label{eq:Trphi_result}
\end{equation}
with $\tilde g_{r\phi}^{(0)} = \Re\,\tilde T_{r\phi}^{(0)}$. Although only the $j=1$ term survives explicitly, the result depends on all levels implicitly through complex conjugated $c^*$, as given in Eq.~\eqref{eq:c1_general}, which requires the quadratic perturbation term. For a~2-level system in the aligned coordinates of Eq.~\eqref{eq:two-level-alignment}, $c=0$ since there are no distant levels and $\H_I^{\{2\}}=0$, recovering $\tilde g_{r\phi}^{(0)}=0$ consistently with Eq.~\eqref{eq:2level_metric_polar}.

\subsection{Local metric diagonalisation}
\label{sec:appendix_metric_diagonalization}
The results above can be simplified with the local metric diagonalisation. For a~general metric tensor, we search for coordinates which locally around the DP diagonalise the metric. This can be expressed by a~parametric modification $\vartheta \equiv \phi+F(r,\phi)$, for some smooth $F$ with $F(0,\phi)=0$, while the radial coordinate is kept, $\rho\equiv r$. Intuitively, this rotates the $r$-axis by an angle that depends on $r$ and $\phi$, eliminating the off-diagonal element. Although $\rho$ and $r$ take the same values, $\partial_\rho$ acts at fixed $\vartheta$, whereas $\partial_r$ acts at fixed $\phi$, so that $g_{\rho\rho}$ generally differs from $g_{rr}$. With $\d r=\d\rho$ and $\d\phi=(\d\vartheta-\partial_rF\,\d\rho)/(1+\partial_\phi F)$, the metric distance transforms into the new coordinates $(\rho,\vartheta)$ as
\begin{equation}
\begin{split}
    \d s^2 &= g_{rr}\d r^2 + 2g_{r\phi} \d r\d \phi+g_{\phi\phi}\d\phi^2\\
    &= \left(g_{rr}-2g_{r\phi}\omega+g_{\phi\phi}\omega^2\right)\d \rho^2 +\frac{2\left(g_{r\phi}-\omega g_{\phi\phi} \right)}{1+\partial_\phi F}\d \rho\,\d\vartheta + \frac{g_{\phi\phi}}{(1+\partial_\phi F)^2}\d\vartheta^2,
\end{split}
\end{equation}
where $\omega\equiv\partial_rF/(1+\partial_\phi F)$. To eliminate the off-diagonal element, we set $\omega=g_{r\phi}/g_{\phi\phi}$. Indeed, such an $F$ exists near the DP: the condition makes $\vartheta$ constant along the curves that cross the circles $r=\mathrm{const.}$ orthogonally, which are well defined, since the metric is continuously differentiable up to $r=0$ and $\tilde g_{\phi\phi}>0$ at a~proper DP, see Eq.~\eqref{eq:limitgphiphi}. Inserting this into the line element, we get
\begin{equation}
    \d s^2 = \underbrace{\left(g_{rr}-\frac{g_{r\phi}^2}{g_{\phi\phi}}\right)}_{g_{\rho\rho}}\d\rho^2 + \underbrace{\frac{g_{\phi\phi}}{(1+\partial_\phi F)^2}}_{g_{\vartheta\vartheta}}\d\vartheta^2,
    \label{eq:diagonalized_metric}
\end{equation}
and the metric tensor in the diagonalised coordinates $g_{\mu\nu}=\mathrm{diag}(g_{\rho\rho},g_{\vartheta\vartheta})$, with the angular component unchanged at the DP, where $\partial_\phi F=0$.

For a~general $N$-level system using Eqs.~\eqref{eq:Tphiphi_result}, \eqref{eq:Trr_result}, \eqref{eq:Trphi_result}, the diagonalised metric component reads
\begin{equation}
    \tilde g_{\rho\rho} = \sum_{j=2}^{N-1} \frac{\left|\braket{\psi_j^{\{0\}}|\H_I^{\{1\}}|\psi_0^{\{0\}}}\right|^2}{\Delta_j^2}+ \left|\braket{\psi_1^{\{0\}}|\psi_0^{\{1\}}}\right|^2\sin^2\left(\arg\left(c\right)-\arg\left(\braket{\psi_1^{\{0\}}|\partial_\phi \H_I^{\{1\}}|\psi_0^{\{0\}}}\right)\right).
\end{equation}
This can be further simplified using the Berry curvature as
\begin{equation}
    \tilde g_{\rho\rho} = \sum_{j=2}^{N-1} \frac{\left|\braket{\psi_j^{\{0\}}|\H_I^{\{1\}}|\psi_0^{\{0\}}}\right|^2}{\Delta_j^2}+ \frac{\tilde V_{r\phi}^2}{4\tilde g_{\phi\phi}}.
\end{equation}
Distant levels contribute the first term (zero for 2-level systems); the degenerate subspace contributes the second, which vanishes for effectively $\PT$-symmetric systems.

\subsection{Geometric circumference of the DP}
\label{sec:appendix_circumference}
Here we derive Eq.~\eqref{eq:c_is_pi} directly from Eqs.~\eqref{eq:limitgphiphi} and~\eqref{eq:circumference_definition}. The Hamiltonian $\H(\ccoord)$ is assumed to be continuously differentiable. Let us call a~chart $\bm x$ \emph{admissible} if the coordinate change $\ccoord\mapsto\bm x$ is continuously differentiable around the DP and has a~non-singular Jacobian at the DP. Let $\bm x=(x_1,x_2)$ be an admissible chart centred at the DP, with polar coordinates $\bm x=r(\cos\phi,\sin\phi)$. Then $\H$ is continuously differentiable in $\bm x$, and its Taylor expansion in $\bm x$, expressed in polar coordinates, gives the two-level subsystem leading term $\hhsub^{\{1\}}(\phi)=\bm f_c\cos\phi+\bm f_s\sin\phi$. Here $\bm f_c,\bm f_s\in\R^3$ are the derivatives of $\bm h_\times$ along $x_1$ and $x_2$ at the DP, and they are linearly independent because the levels split linearly in every direction.\footnote{Aligned coordinates~\eqref{eq:two-level-alignment} correspond to the special case $\bm f_c=(1,0,0)$, $\bm f_s=(0,0,1)$.} Inserting $\hhsub^{\{1\}}$ into Eq.~\eqref{eq:limitgphiphi} gives
\begin{equation}
    \tilde g_{\phi\phi}^{(n)}=\frac{1}{4}\frac{|\bm f_c\times\bm f_s|^2}{\left(|\bm f_c|^2\cos^2\phi+|\bm f_s|^2\sin^2\phi+(\bm f_c\cdot\bm f_s)\sin 2\phi\right)^2}.
\end{equation}
Because it depends only on $\xi_c\equiv|\bm f_c|^2$, $\xi_{cs}\equiv\bm f_c\cdot\bm f_s$ and $\xi_s\equiv|\bm f_s|^2$, with $\xi_c\xi_s-\xi_{cs}^2=|\bm f_c\times\bm f_s|^2$, it becomes
\begin{equation}
    \sqrt{\tilde g_{\phi\phi}^{(n)}}=\frac{1}{2}\sqrt{\frac{\xi_c\xi_s-\xi_{cs}^2}{\left(\xi_c\cos^2\phi+\xi_s\sin^2\phi+\xi_{cs}\sin 2\phi\right)^2}},
\end{equation}
where $\xi_c>0$, $\xi_s>0$ and $\xi_c\xi_s>\xi_{cs}^2$ follow from the linear independence of $\bm f_c$ and $\bm f_s$. Inserting this into Eq.~\eqref{eq:circumference_definition}, where the limit can be taken under the integral since Eq.~\eqref{eq:limitgphiphi} holds uniformly in $\phi$, and performing the integration (for example in Mathematica), we directly get $\CC=\pi$. The result does not depend on $\bm f_c$ and $\bm f_s$, so it holds in every admissible chart.

\section{Linear 3-level system}
\label{sec:appendix_3level_details}
The algebraic structure of the Lie group generators is characterised by the commutation and anticommutation relations
\begin{align}
    [\hat\lambda_a, \hat\lambda_b] = 2i f_{abc}\,\hat\lambda_c,\quad 
    \{\hat\lambda_a, \hat\lambda_b\} = \tfrac{4}{N}\delta_{ab}\Id + 2d_{abc}\,\hat\lambda_c.
\end{align}
$f_{abc}$ are totally antisymmetric and $d_{abc}$ are totally symmetric structure constants
\begin{equation}
    f_{abc} \equiv -\tfrac{i}{4}\Tr([\hat\lambda_a,\hat\lambda_b]\hat\lambda_c), \qquad
    d_{abc} \equiv \tfrac{1}{4}\Tr(\{\hat\lambda_a,\hat\lambda_b\}\hat\lambda_c),
    \label{eq:structure_constants}
\end{equation}
defining the cross product $(\bm a \times_L \bm b)_i \equiv f_{ijk} a_j b_k$ and star product $(\bm a \star \bm b)_i \equiv d_{ijk} a_j b_k$ in $\R^{N^2-1}$.

For 3-level systems, we choose the Gell-Mann matrices $\{\hat\lambda_k\}_{k=1}^8$ as generators of the $su(3)$ algebra \cite{Liu07}. They are explicitly
\begin{align}
\lambda_1 &= \begin{pmatrix}
0 & 1 & 0 \\
1 & 0 & 0 \\
0 & 0 & 0
\end{pmatrix},\;\;
\lambda_2 = \begin{pmatrix}
0 & -i & 0 \\
i & 0 & 0 \\
0 & 0 & 0
\end{pmatrix},\;\;
\lambda_3 = \begin{pmatrix}
1 & 0 & 0 \\
0 & -1 & 0 \\
0 & 0 & 0
\end{pmatrix},\;\;
\lambda_4 = \begin{pmatrix}
0 & 0 & 1 \\
0 & 0 & 0 \\
1 & 0 & 0
\end{pmatrix},\;\;\\
\lambda_5 &= \begin{pmatrix}
0 & 0 & -i \\
0 & 0 & 0 \\
i & 0 & 0
\end{pmatrix},\;\;
\lambda_6 = \begin{pmatrix}
0 & 0 & 0 \\
0 & 0 & 1 \\
0 & 1 & 0
\end{pmatrix},\;\;
\lambda_7 = \begin{pmatrix}
0 & 0 & 0 \\
0 & 0 & -i \\
0 & i & 0
\end{pmatrix},\;\;
\lambda_8 = \frac{1}{\sqrt{3}} \begin{pmatrix}
1 & 0 & 0 \\
0 & 1 & 0 \\
0 & 0 & -2
\end{pmatrix}.
\end{align}

The Bloch vector for the $n$th eigenstate in $N=3$ system is
\begin{equation}
    \bm b^{(n)}(\ccoord) = 2\frac{E_n(\ccoord) \h(\ccoord) + \h(\ccoord)\star \h(\ccoord)}{3E_n(\ccoord)^2 - \frac{1}{2}\sum_{k=0}^{2} E_k(\ccoord)^2}.
    \label{eq:Bloch_3level}
\end{equation}
To describe the DP of a~general 3-level system, we need the second-order expansion in Eq.~\eqref{eq:H_expansion_DP}. Because such a~system requires 16 parameters, for simplicity we restrict the analysis to the first-order expansion in the spectrum~\eqref{eq:spectrum_linearization}, corresponding to the Hamiltonian
\begin{equation}
    \bm h(r,\phi) = \sqrt{3}(0,\dots,0,-1)^T+r (f_1,\dots,f_7,\sqrt{3}f_8)^T,
    \label{eq:3level_hamiltonian_vector}
\end{equation}
for all functions $\phi$-dependent, which has a~matrix representation
\begin{equation}
    \H(r,\phi) = \begin{pmatrix}
        -1&0&0 \\
        0&-1&0 \\
        0&0&2
    \end{pmatrix}+r
    \begin{pmatrix}
        f_3 + f_8 & f_1 - if_2 & f_4 - if_5 \\
        f_1 + if_2 & -f_3 + f_8 & f_6 - if_7 \\
        f_4 + if_5 & f_6 + if_7 & - 2f_8
    \end{pmatrix}.
    \label{eq:H_3level}
\end{equation}
The angular component of the metric tensor depends only on the 2-level subsystem vector $\hhsub(\phi) \equiv (f_1(\phi), f_2(\phi), f_3(\phi))^T$, as in Eq.~\eqref{eq:hsubvector}. The radial and off-diagonal components, by contrast, receive contributions from coupling to the third level and do not vanish as they do for the 2-level system. In the limit $r\rightarrow 0$, the metric tensor is
\begin{align}
    \tilde g_{rr}^{(0)} &= \frac{(A^2+B^2+C^2+D^2)||\hhsub||^2+\frac{1}{4}||\hhsub\times \bm S||^2}{36||\hhsub||^4}, \label{eq:app_limitgrr}\\
    \tilde g_{r\phi}^{(0)} &= -\frac{\bm S\cdot \bm v}{24 ||\hhsub||^4},\label{eq:app_limitgrphi}\\
    \tilde g_{\phi\phi}^{(0)} &= \frac{1}{4}\left\|\frac{\hhsub}{|\hhsub|} \times \partial_\phi\left(\frac{\hhsub}{|\hhsub|}\right)\right\|^2,\label{eq:app_limitgphiphi}
\end{align}
for complex combinations $Z_1 \equiv f_4 + i f_5$ and $Z_2 \equiv f_6 + i f_7$, $\bm v \equiv -\hhsub \times (\hhsub \times \hhsub')$,
\begin{align}
    A &\equiv f_4 (f_3-||\hhsub||) + f_1 f_6 - f_2 f_7, \\
    B &\equiv f_5 (f_3-||\hhsub||) + f_1 f_7 + f_2 f_6, \\
    C &\equiv f_6 (f_3+||\hhsub||) - f_1 f_4 - f_2 f_5, \\
    D &\equiv f_7 (f_3+||\hhsub||) - f_1 f_5 + f_2 f_4,
\end{align}
and the vector $\bm S$
\begin{equation}
    \bm S \equiv \begin{pmatrix}
        2 \Re(Z_1 Z_2^*) \\
        2 \Im(Z_1 Z_2^*) \\
        |Z_1|^2 - |Z_2|^2
    \end{pmatrix}.
\end{equation}

\subsection{Aligned coordinates}
\label{sec:aligned_coordinates}
We can align the coordinates of the two-level subspace participating in the degeneracy using Eq.~\eqref{eq:two-level-alignment}, where $h^i$ now denote the Gell-Mann matrices, and we align only the two-level subspace locally near the DP. The metric components simplify to
\begin{align}
    \tilde g_{rr}^{(0)} &= \frac{4(A^2+B^2+C^2+D^2)+||(\cos\phi,0,\sin\phi)^T\times \bm S||^2}{144},
    \label{eq:limitgrr}\\
    \tilde g_{r\phi}^{(0)} &= -\frac{\bm S\cdot (-\sin\phi,0,\cos\phi)^T}{24},
    \label{eq:limitgrphi}\\
    \tilde g_{\phi\phi}^{(0)}&= \frac{1}{4}.
    \label{eq:app_limitgphiphi_aligned}
\end{align}
The metric can be further simplified for systems with effective $\PT$ symmetry ($f_5=f_7=0$), leading to 
\begin{equation}
       \H(r,\phi) = 
\begin{pmatrix}
 -1+r (\sin \phi +f_8) & r \cos\phi & r f_4 \\
 r \cos\phi & -1-r (\sin \phi-f_8 )& r f_6 \\
 r f_4 & r f_6 & 2-2 r f_8 \\
\end{pmatrix},
\label{eq:H_real}
\end{equation}
using transformation
\begin{equation}
\begin{pmatrix}
\beta\\
\eta
\end{pmatrix}
\equiv 
\begin{pmatrix}
\cos\!\tfrac{\phi}{2} & -\sin\!\tfrac{\phi}{2}\\[6pt]
\sin\!\tfrac{\phi}{2} & \cos\!\tfrac{\phi}{2}
\end{pmatrix}
\begin{pmatrix}
f_4\\[3pt]
f_6
\end{pmatrix}.
\end{equation}
The metric then takes the form
\begin{equation}
    \tilde g_{rr}^{(0)} = \frac{(\eta-\beta)^2\big((\beta+\eta)^2+8\big)}{144}, \quad   \tilde g_{r\phi}^{(0)} = \frac{\eta^2-\beta^2}{24},\quad\tilde g_{\phi\phi}^{(0)} = \frac{1}{4}.
    \label{eq:limitgphiphiRealAligned}
\end{equation}

In the directions $\phi$ in which $g_{rr}=g_{r\phi}=0$, this limiting metric coincides with that of the two-level system, Eq.~\eqref{eq:2level_metric_polar}. This happens for $\beta=\eta$, or equivalently, if the couplings $f_4$ and $f_6$ of the degenerate pair to the third level satisfy
\begin{equation}
    \frac{f_6}{f_4}= \frac{2}{\tan{\frac{\phi }{2}}+1}-1.
    \label{eq:h4h6_zero_condition}
\end{equation}
The limiting metric~\eqref{eq:limitgphiphiRealAligned} has the determinant $\det\tilde g^{(0)}=(\eta-\beta)^2/72$, where $(\eta-\beta)/\sqrt{2}$ is, up to sign, the first-order coupling $\braket{\psi_2^{\{0\}}|\H_I^{\{1\}}|\psi_0^{\{0\}}}$ of the lower DP state to the third level. The limiting metric is thus degenerate in the directions in which this coupling vanishes. For $\H$ differentiable at the DP, these directions are isolated unless $f_4=f_6=0$, in which case the third level decouples at first order and the DP is effectively that of a~two-level system. As this limit is approached, e.g.\ by scaling $f_4$ and $f_6$ by $\varepsilon\to 0$, $\det\tilde g^{(0)}\propto\varepsilon^2$ vanishes on the whole horizon, while $\tilde g^{(0)}_{\phi\phi}=1/4$, the circumference $\pi$, the glueing and the continuity of the metric remain unaffected. Only the geodesic equation, which needs the inverse metric, becomes singular on the whole horizon in the limit.

The spin-1 model~\eqref{eq:3-level_H} is far from this decoupled limit. In polar coordinates $(r,\phi)$ centred at either of its DPs, in aligned coordinates oriented such that $\phi=\pi/2$ points towards the other DP, in the eigenbasis of $\H$ at this DP with $\ket{m_s=0}$ as the second basis vector, and after an affine change of the energy scale, the model takes exactly the form~\eqref{eq:H_real}, at both DPs with the same functions $f_4=2e\sin\phi/\sqrt{D^2-e^2}\approx 0.35\sin\phi$, $f_6=\sqrt{(D+e)/(D-e)}\cos\phi\approx 1.19\cos\phi$ and $f_8=\sin\phi$, where the approximate values are calculated for the parameters of Sec.~\ref{sec:spin-1-system}. The couplings are therefore not small, and $\eta-\beta$ vanishes on $\M^0$ only for $\phi=\pi/2$, i.e.\ in the direction of the other DP, along the zero-determinant line $B_x=0$.

\subsection{Perturbation theory}
\label{sec:appendix_SPT_3-level}
\begin{figure}
    \centering
    \includegraphics{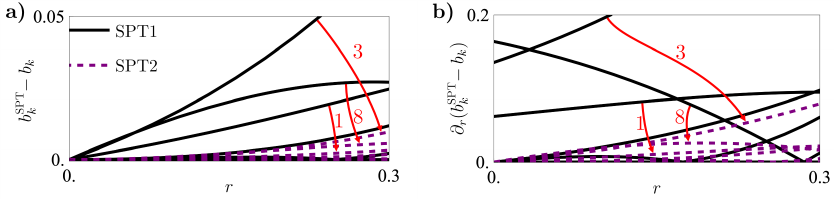}
    \caption{Demonstration of a~perturbation theory for a~linear system~\eqref{eq:H_real} with test functions $f_4=\sin\phi\cos\phi$, $f_6=-3\cos\phi$, $f_8=\sin\phi$ along fixed $\phi=2$ direction. (a) The difference between exact and perturbation theory (SPT) $\bm b_0(r)$ vector elements. (b) The difference of the Bloch vector's radial derivative. First-order perturbation theory (SPT1, solid black) and second-order (SPT2, dashed purple). Indices of $b_k$ with the largest difference are marked, $k\in\{1, 3, 8\}$, along with red arrows marking the change from first- to second-order SPT.}
    \label{fig:grrProblem}
\end{figure}

We demonstrate the stationary perturbation theory (SPT) from Appendix \ref{sec:appendix_SPT} for the 3-level system~\eqref{eq:H_real}. The spectrum up to the second order (SPT2) is
\begin{equation}
    \begin{split}
        E_0^{\mathrm{SPT2}}(r, \phi) &= -1+r (f_8(\phi )-1)+\frac{1}{6} r^2 (\sin \phi -1) (f_4(\phi )-f_6(\phi ) (\tan \phi +\sec \phi ))^2,\\
        E_1^{\mathrm{SPT2}}(r, \phi) &= -1+r (f_8(\phi )+1)-\frac{r^2 (f_4(\phi ) (\sin \phi +1)+f_6(\phi ) \cos \phi )^2}{6 (\sin \phi +1)}\\
        E_2^{\mathrm{SPT2}}(r, \phi)&=-E_0^{\mathrm{SPT2}}(r, \phi)-E_1^{\mathrm{SPT2}}(r, \phi).
    \end{split}
\end{equation}
In the limit $r\to 0$, $\bm b$ needed to compute the metric tensor in Eq.~\eqref{eq:geometric_tensor_Bloch} takes the indeterminate form $"0/0"$. First-order perturbation theory correctly reproduces $\lim_{r\to 0}\partial_\phi \bm b_0$, but yields an incorrect $\lim_{r\to 0}\partial_r \bm b_0$, see Fig.~\ref{fig:grrProblem}. As a~result, the angular metric tensor element $\tilde{g}_{\phi\phi}$ is correctly reproduced by SPT1, but other elements converge to incorrect values. Second-order perturbation theory corrects this, and $\lim_{r\to 0}\partial_r \bm b_0=\lim_{r\to 0}\partial_r \bm b_0^{\mathrm{SPT2}}$.

\section{Adjugate matrix method for eigenvector calculation}
\label{sec:appendix_AMM}

For an $N$-dimensional Hamiltonian $\H(\ccoord)$, the adjugate matrix method can determine eigenvectors with a~fixed gauge. The gauge often cannot be fixed globally; consequently, this method fails on some subspace of parametric space. This failing subspace determines the nodal lines discussed in Sec.~\ref{sec:nodal_lines}.

We first define (omitting the $\ccoord$ dependence)
\begin{equation}
    M_k\equiv \H-E_k \Id,
\end{equation} 
for which $\det(M_k)=0$ if $E_k$ is an eigenvalue. We assume that the eigenvalue $E_k$ is non-degenerate, so that $\mathrm{rank}(M_k)=N-1$ (equivalently, $\mathrm{rank}(M_k)>N-2$). This condition excludes DPs, but our formalism approaches them only as limits, so non-degeneracy holds in the neighbourhood. Then 
\begin{equation}
    \adj(M_k)M_k=\det(M_k)\Id = 0 = M_k \adj(M_k).
\end{equation} 
We define a~vector 
\begin{equation}
    \bm\Psi_k \equiv \adj(M_k)\bm w \text{ for any } \bm w\in \mathbb C^N\backslash\{0\},
\end{equation}
for which $M_k \bm\Psi_k = M_k \adj(M_k) \bm w = 0$ implying $\bm\Psi_k \in \mathrm{Ker} M_k$, proving $\bm\Psi_k$ is an eigenvector of $\H$ corresponding to eigenvalue $E_k$.

For small dimensions:
\begin{itemize}
        \item $N=1$: $\adj M=[1]$,
        \item $N=2$: $\adj \begin{pmatrix}
            a_1&a_2\\ b_1&b_2
        \end{pmatrix}=\begin{pmatrix}
            b_2 & -a_2\\ -b_1& a_1
        \end{pmatrix}$,
        \item $N=3$: $\adj \begin{pmatrix}
            a_1& a_2 & a_3\\
            b_1& b_2 & b_3\\
            c_1& c_2 & c_3\\
        \end{pmatrix}=\begin{pmatrix}
            \left|\begin{matrix}
                b_2&b_3\\c_2&c_3
            \end{matrix}\right|& -\left|\begin{matrix}
                b_1&b_3\\c_1&c_3
            \end{matrix}\right| & .\\
            .& . & .\\
            .& . & .\\
    \end{pmatrix}^T$. Up to the transposition, this is the cofactor matrix.

\end{itemize}

This means that the columns of the adjugate matrix $\adj M_k$ are the eigenvectors corresponding to eigenvalue $E_k$, and $\bm w$ represents the gauge choice. The eigenvectors in the columns are related by a~complex phase determined by $\bm w$. For a~real Hamiltonian and a~real gauge choice $\bm w \in \mathbb{R}^N$, the parameterisation is redundant: $\bm w$ has $N$ real components, but only $N-1$ real parameters matter, since the overall magnitude is irrelevant.

This eigenvector calculation process is sometimes referred to as the \emph{adjugate matrix method}, see e.g. \cite{Denton22}, and it correctly fails along the nodal lines (returning the zero vector) and yields a~consistent gauge everywhere else.

\section{Unstable lines for geodesic passages}
\label{sec:appendix_singular_lines}
In this section, we consider a~simplified model of a~zero-determinant line from Section~\ref{sec:singular_lines}. We assume a~state manifold over parametric space $\ccoord\equiv (x,z)\in\R^2$ with metric tensor 
\begin{equation}
g(x,z) \equiv \begin{pmatrix} g_{xx}(z)& 0 \\ 0 & g_{zz}(x,z)\end{pmatrix} = \begin{pmatrix} z^{2p}& 0 \\ 0 & g_{zz}(x,z)\end{pmatrix},
\end{equation}
for some $p>0$. Such a~toy model has a~zero-determinant line at $z=0$ since the metric element $g_{xx}\rightarrow 0$ as $z\rightarrow 0$, and $\det(g)\rightarrow 0$. The Christoffel symbols~\eqref{eq:geodesicEq} are explicitly
\begin{align}
    \Gamma^x_{\;\;zz} &= -\frac{g_{zz,x}}{2z^{2p}}, \;\;\;\;\quad \Gamma^x_{\;\;xz} = \frac{p}{z}, \quad\quad\; \Gamma^x_{\;\;xx} = 0,\\
    \Gamma^z_{\;\;xx} &= -p \frac{z^{2p-1}}{g_{zz}}, \quad \Gamma^z_{\;\;xz} = \frac{g_{zz,x}}{2g_{zz}}, \quad \Gamma^z_{\;\;zz} = \frac{g_{zz,z}}{2g_{zz}}.
\end{align}
For convenience, we further assume that near the zero-determinant line $g_{zz}(x,z)=g_{zz}(z)$.
The geodesic equation
\begin{equation}
    \ddot{\alpha}^\mu + \Gamma^\mu_{\;\;\kappa\lambda}\dot{\alpha}^\kappa\dot{\alpha}^\lambda = 0
\end{equation}
then simplifies to
\begin{align}
    \ddot{x} + \frac{2p}{z}\dot x \dot z &= 0,\\
    \ddot{z} - p \frac{z^{2p-1}}{g_{zz}(z)}\dot x \dot x + \frac{1}{2}\frac{g_{zz,z}}{g_{zz}}\dot z \dot z &= 0.
\end{align}
The first equation is equivalent to 
\begin{equation}
    \frac{d}{dt}\left(z^{2p}\dot x(t)\right) = 0 \Leftrightarrow z^{2p}\dot x(t) = c;\quad c\in\R
\end{equation}
This demonstrates that as $z\rightarrow 0$, $\dot x(t) \rightarrow \infty$ for any geodesic with $c \neq 0$, meaning as a~geodesic approaches the zero-determinant line, it deflects in the $x$-direction. It also shows that the only possibility for a~geodesic to cross the line is to have $\dot x = 0$, meaning the geodesic must arrive perpendicularly to the zero-determinant line. 

The $z$-component of the geodesic equation cannot be explicitly solved until $g_{zz}(z)$ is specified, as it is more complicated. For example, for general $g_{zz}(z)$, the equation reads:
\begin{equation}
    \ddot z(t) + \frac{g_{zz,z}(z(t))}{2g_{zz}(z(t))} \dot z(t) \dot z(t)- \frac{p c^2 z(t)^{-2p-1}}{g_{zz}(z(t))} = 0.
\end{equation}

\newpage

\end{appendix}

\bibliography{SciPost_BiBTeX.bib}

\end{document}